\documentclass[reqno,a4paper]{amsart}
\usepackage{hyperref}
\usepackage{graphicx}
\usepackage{multicol}
\usepackage{color}
\usepackage{caption}
\usepackage{setspace}
\usepackage{amssymb,amsmath}  
\numberwithin{equation}{section}   
\usepackage{amsfonts,amsrefs}
\usepackage{amsthm} 
\usepackage{bm}    
\usepackage{bbm}      
\usepackage{tabularx}
\usepackage{enumerate}
\usepackage{algorithmic}
\usepackage{lscape}
\makeatletter
\def\verbatim@font{\linespread{1}\normalfont\ttfamily}
\makeatother
\usepackage{pgfpages}
\usepackage{changepage}
\usepackage{fancyhdr,a4wide}
\usepackage{forest}
\usepackage{listings}
\usepackage{placeins}

\usepackage[utf8]{inputenc}
\usepackage[T1]{fontenc}
\usepackage{amsmath}
\usepackage{tensor}
\usepackage{amssymb}
\usepackage{graphicx}
\usepackage{subfigure}
\usepackage{mathrsfs}
\usepackage{microtype}
\usepackage{rotating}
\usepackage{xspace}
\usepackage{mathtools}
\usepackage{tikz,pgfplots}
\usepackage{comment}
\usepackage{MnSymbol}
\pgfplotsset{compat=1.9}
\usetikzlibrary{fit}
\tikzset{every label/.style={font=\footnotesize,inner sep=1.5pt}}
\usepackage{dutchcal} 

\theoremstyle{plain}
\newtheorem{thm}{Theorem}[section]

\newtheorem*{thm*}{Theorem}

\theoremstyle{definition}

\theoremstyle{remark}
\newtheorem{rem}[thm]{Remark}

\input{Tdef.def}

\allowdisplaybreaks
\def\jf#1{\relax}

\title[Mixmaster Fluids Near the Big Bang]{Mixmaster Fluids Near the Big Bang}

\author[E.~Marshall]{Elliot Marshall}
\address{School of Mathematics\\
9 Rainforest Walk\\
Monash University, VIC 3800\\ Australia}
\email{elliot.marshall@monash.edu}

\begin{document}

\begin{abstract}
    We numerically study the approach to the singularity in $\mathbb{T}^{2}$-symmetric cosmological spacetimes containing a non-stiff perfect fluid satisfying a linear equation of state $p=K\rho$, $K \in [0,1)$. Near the singularity, the dynamics are found to be local and oscillatory. In particular, our results show, for the first time, that the fluid velocity in inhomogeneous cosmologies develops mixmaster-esque oscillations consistent with the generalised BKL conjecture of Uggla et al. Moreover, we find these fluid oscillations are responsible for the development of local inhomogeneities in the matter density of the early universe.
\end{abstract}

\maketitle

\section{Introduction}
The Penrose-Hawking singularity theorems \cite{HawkingEllis:1973} guarantee that a wide class of cosmological models are past geodesically incomplete. However, these theorems do not provide any information about the nature of these `big bang' singularities. A longstanding problem in mathematical cosmology, therefore, has been to understand the dynamical behaviour of solutions to the Einstein equations near singularities. In the seminal work of Belinski\v{\i}, Khalatnikov, and Lifschitz (BKL) \cite{belinskii1970}, it was conjectured that the initial singularity is \textit{spacelike} and \textit{local}. Roughly speaking, `local' means that, near the big bang, solutions are well-approximated by the spatially homogeneous equations obtained by setting spatial derivatives to zero in the Einstein equations. That is to say, solutions near the big bang are expected to display ODE-dominated behaviour. A generalised and rigorously stated version of the BKL conjecture has subsequently been given by Uggla et al.\ \cite{UEWE:2003}. \newline \par

The nature of the local dynamics near the singularity depends on the matter source of the system. For stiff matter, such as a massless scalar field or stiff fluid ($p=\rho$), the approach to the big bang is expected to be \textit{quiescent}, where the solution converges, pointwise, to a \textit{Kasner} spacetime. The Kasner spacetimes are a one-parameter family of solutions to the vacuum Einstein equations defined by
\begin{align}
\label{eqn:Kasner_Metric}
    g = -dt\otimes dt + \sum_{\Omega=1}^{3}t^{2P_{\Omega}}dx^{\Omega}\otimes dx^{\Omega}.
\end{align}
where the constants $P_{\Omega}$ ($\Omega =1,2,3$) are known as the \textit{Kasner exponents} and satisfy the conditions
\begin{align}
\label{eqn:Kasner_Conditions}
    \sum_{\Omega = 1}^{3}P_{\Omega} = 1, \quad \sum_{\Omega = 1}^{3}P_{\Omega}^{2} = 1.
\end{align}
In recent years, numerous rigorous stability results have been established on the formation of quiescent big bang singularities in inhomogeneous spacetimes, see for example \cites{andersson2001,RodnianskiSpeck:2018c,Fournodavlos_et_al:2023,Groeniger_et_al:2023,BeyerOliynyk:2023,BeyerOliynyk:2024,BOZ:2025}. We also note that quiescent behaviour was observed in numerical simulations of stiff fluid cosmologies by Curtis and Garfinkle \cite{CurtisGarfinkle:2005}. \newline \par

On the other hand, for vacuum and non-stiff matter, BKL conjectured that the local dynamics are \textit{oscillatory}. In these models, the solution chaotically oscillates between different Kasner spacetimes\footnote{That is to say, at each point in space the Kasner exponents oscillate between different values.} at each point in space as the singularity is approached. This is often called \textit{mixmaster} dynamics. Rigorous results in this setting have thus far been limited to spatially homogeneous spacetimes \cites{Ringstrom:2001,HeinzleUggla:2009,Brehm:2016}, although there is an extensive body of numerical work for vacuum spacetimes which supports the BKL picture, see for example \cites{BergerIsenbergWeaver:2001,Garfinkle:2004,Andersson_et_al:2005}. There has been comparatively little research, however, into the dynamics of inhomogeneous cosmologies containing non-stiff matter near the big bang\footnote{We note that the work of Weaver et al.\ \cite{Weaver_et_al:1998} studies inhomogeneous models containing a pure magnetic field. However, in this case the matter is non-dynamical and simply contributes a constant term to the Einstein equations.}. \newline \par  

The purpose of this article is to report on numerical simulations of the approach to the big bang singularity in $\mathbb{T}^{2}$-symmetric cosmological spacetimes\footnote{See Section \ref{sec:Einstein_Euler_Eqns} for further details about this symmetry class.} containing a non-stiff perfect fluid. The stress-energy tensor of the fluid is given by 
\begin{align*}
    T^{\mu\nu} = (\rho + p)v^{\mu}v^{\nu} + pg^{\mu\nu}, \quad v^{\nu}v_{\nu} = -1,
\end{align*}
where $v^{\mu}$ is the fluid four-velocity, $\rho$ is the fluid's proper energy density, and $p$ is the pressure. We assume that $\rho$ and $p$ are related by a linear equation of state
\begin{align*}
    p = K\rho 
\end{align*}
where $K=c_{s}^{2}$ is the square of the sound speed. In this article we focus on the case $K \in [0,1)$, which corresponds to the non-stiff fluid\footnote{As mentioned above, stiff fluid $(K=1)$ models are quiescent, see for example \cite{CurtisGarfinkle:2005}.}.  \newline \par

Although it is commonly claimed that `matter does not matter' near the big bang, the presence of a non-stiff fluid is, in fact, expected to change the past attractor of generic cosmological spacetimes through the addition of so-called \textit{tilt transitions} \cites{Hewitt_et_al:2001,UEWE:2003}. The concept of fluid tilt is defined relative to a fixed timelike vector $u = u^{\mu}\partial_{\mu}$. We say a fluid is \textit{orthogonal} with respect to $u$ if its four-velocity $v=v^{\mu}\partial_{\mu}$ is aligned with $u$ and \textit{tilted} otherwise. For a general spacetime without symmetries, there is no preferred choice of $u$. However, for spatially homogeneous models, there exists a distinguished foliation by hypersurfaces of homogeneity. In this case, it is natural to take $u$ to be the normal vector of the foliation. Moreover, for solutions of the field equations which are well approximated (pointwise) by spatially homogeneous models near the big bang, we can talk of a fluid being \textit{asymptotically} orthogonal or tilted at each individual point in space. \newline \par 

In practice, following \cite{UEWE:2003}, we define the fluid tilt relative to an orthonormal frame $\{e_{i}\}$, $i=0,1,2,3$, where $e_{0} = u = \alpha^{-1}\partial_{t}$ is assumed to be hypersurface orthogonal and $\alpha$ is the lapse of the foliation. With respect to $e_{0}$, the fluid four-velocity is then decomposed as
\begin{align}
\label{eqn:tilt_decomposition_intro}
    v^{a} = \Gamma(e_{0} + \nu^{a}), \quad \Gamma = \frac{1}{\sqrt{1-|\nu|^{2}}}, \quad |\nu|^{2} = \nu^{A}\nu_{A},
\end{align}
where $|\nu|<1$. The three-velocity $\nu^{A}$ is orthogonal to $e_{0}$ and determines the tilt of the fluid. That is to say, the fluid is orthogonal when $|\nu| = 0$ and tilted otherwise. As $|\nu| \nearrow 1$, the fluid approaches a null vector and is said to be \textit{extremely tilted}. A tilt transition is where the fluid velocity goes from an orthogonal to an extremely tilted state (and vice versa). Thus, for cosmological models containing a non-stiff fluid, we expect oscillatory behaviour in both the gravitational variables \textit{and} the fluid velocity.  \newline \par 

While tilt transitions were first conjectured to occur in the work of Hewitt et al.\ on tilted Bianchi II cosmologies \cite{Hewitt_et_al:2001}, there have been no rigorous results about the asymptotics of oscillatory fluid models to date, even in the spatially homogeneous setting. Additionally, in the only previous numerical study of non-stiff, fluid-filled, inhomogeneous models \cite{Lim_Thesis:2004}, no tilt transitions were observed. However, these simulations were relatively short and only considered the case where the fluid had one non-zero spatial component $\nu^{A} = \nu\delta^{A}_{1}$. We also note there has been extensive work on the future asymptotics of tilted fluid models. In particular, Coley, Hervik, and collaborators have analysed the influence of tilted fluids for many types of Bianchi cosmologies \cites{Coley_et_al:2006a,Coley_et_al:2006b,ColeyHervik:2004,ColeyHervik:2008,Hervik_et_al:2006,Hervik_et_al:2007a,Hervik_et_al:2010,Hervik_et_al:2007b,ColeyHervik:2005}. Similarly, the non-linear future stability of certain tilted Bianchi models has recently been established in \cites{Fournodavlos_et_al:2024,FMO:2025}. However, in all of these examples the fluid approaches either an orthogonal or extremely tilted state towards future timelike infinity and tilt transitions are \textit{not} present. \newline \par

In this article, we employ a new numerical approach based on \textit{path-conservative} finite volume schemes\footnote{See Section \ref{sec:Numerical_Setup} for details of our numerical implementation.}. This method allows us to stably model discontinuities in the fluid while directly evolving the primitive fluid variables, which have better asymptotic behaviour than the standard conserved variables. Importantly, by using this path-conservative approach, we have been able to evolve inhomogeneous solutions which contain multiple tilt transitions in the fluid velocity for the first time. In particular, our results provide convincing support for the generalised BKL conjecture of Uggla et al.\ \cite{UEWE:2003} in fluid-filled, inhomogeneous cosmological spacetimes. \newline \par 

 The remainder of the article is structured as follows: first, in Section \ref{sec:Einstein_Euler_Eqns} we introduce the form of the $\mathbb{T}^{2}$-symmetric Einstein-Euler equations which we numerically solve. This is followed, in Section \ref{sec:Homogeneous_Asymptotics}, by a discussion of the expected asymptotic behaviour near the big bang and a derivation of `trigger' conditions for tilt transitions in the fluid velocity. Next, in Section \ref{sec:Numerical_Setup}, we discuss our numerical implementation, choice of initial data, and code tests. In Section \ref{sec:Numerical_Results}, we  present the results of our simulations and compare them to the generalised BKL framework of \cite{UEWE:2003}. We conclude with an outline of future work.

\subsubsection{Indexing Conventions}
Our indexing conventions are as follows: lower case Greek letters, e.g. $\mu,\nu,\kappa$, will label spacetime coordinate indices that run from $0$ to $3$ while upper case Greek letters, e.g. $\Gamma,\Omega,\Sigma$, will label spatial coordinate indices that run from $1$ to $3$. We also employ orthonormal frames $e_{a} = e^{\mu}_{a}\del_{\mu}$. Lower case Latin letters, e.g. $a$, $b$, $c$, that run from $0$ to $3$ will label spacetime frame indices while spatial frame indices will be labelled by upper case
Latin letters, e.g. $A$, $B$, $C$, and run from $1$ to $3$. Lower case calligraphic Latin indices, e.g. $\mathcal{a}$, $\mathcal{b}$, denote frame indices
which run from 2 to 3.

\subsubsection{Index Operations}
The \textit{symmetrisation}, \textit{anti-symmetrisation}, and \textit{symmetric trace-free} operations on pairs of spatial frame indices are defined by
\begin{gather*}
    \Sigma_{(AB)} = \frac{1}{2}(\Sigma_{AB} + \Sigma_{BA}), \quad \Sigma_{[AB]} = \frac{1}{2}(\Sigma_{AB}-\Sigma_{BA}),\intertext{and}
    \Sigma_{<AB>} = \Sigma_{(AB)}- \frac{1}{3}\delta^{CD}\Sigma_{CD}\delta_{AB},
\end{gather*}
respectively.

\section{\texorpdfstring{$\beta$}{}-Normalised Einstein-Euler Equations}
\label{sec:Einstein_Euler_Eqns}
In this article, we work with a symmetry reduced version of the orthonormal frame formalism for the Einstein equations. This approach was pioneered in a series of works by Ellis \& MacCallum \cites{Ellis:1967, EllisMaccallum:1969,MacCallum:1973} and  subsequently extended by van Elst and Uggla \cite{ElstUggla:1997} and Uggla et al. \cite{UEWE:2003}. One of the key advantages of this formulation is that the spatially homogeneous Einstein equations reduce to a system of autonomous ODEs, which allows for the application of powerful techniques from the study of dynamical systems. See \cites{Ringstrom:2001, EllisKing:1973, UEWE:2003, WainwrightHsu:1989, Garfinkle:2004,Andersson_et_al:2005,BOZ:2025,FMO:2025} for a selection of analytic and numerical works using this formalism. \newline \par 

We consider cosmological models where a $\mathbb{T}^{2}$ isometry group acts on spacelike hypersurfaces with $\mathbb{T}^{3}$ topology. Under this assumption, the Einstein-Euler equations reduce to a $1+1$ dimensional system\footnote{That is, all variables depend only on $(t,x)$.}. For $\mathbb{T}^{2}$-symmetric spacetimes, it is natural to work with the $\beta$-normalised frame equations, first derived in \cite{Elst_et_al:2001}. In this case, the frame variables are normalised by the area expansion rate of the $\mathbb{T}^{2}$-symmetry orbits and the evolution and constraint equations take a particularly simple form. Importantly, these expansion normalised variables remain \textit{bounded} near the singularity. Indeed, the use of expansion-normalised variables appears to be essential in order to achieve stable simulations near the big bang, see for example \cites{Garfinkle:2004,Garfinkle:2007,Andersson_et_al:2005,GarfinklePretorius:2020,ColeyLim:2012,Lim:2009}. \newline \par

A derivation of the $\beta$-normalised equations is given in the thesis of Lim \cite{Lim_Thesis:2004} and the work of van Elst et al.\ \cite{Elst_et_al:2001}. We work in the timelike areal gauge which has been widely used in previous numerical studies of Gowdy and $\mathbb{T}^{2}$-symmetric cosmologies, see for example \cites{BergerIsenbergWeaver:2001,LeFlochRendall:2011,BMO:2023,BMO:2024,Marshall:2025,BergerGarfinkle:1998, ColeyLim:2023}. Our spatial gauge is determined by requiring the frame vector $e_{2}$ is parallel to a Killing vector (cf. \cite{Elst_et_al:2001}). 
Note, that this is the same gauge conditions as in the standard `Gowdy' parametrisation of $\mathbb{T}^{2}$-symmetric spacetimes, cf. \cite{BergerIsenbergWeaver:2001}. \newline \par

For our gauge choices, the evolution equations become 
\begin{align}
\label{eqn:SigmaMinus_ArealGauge}
\del_{t}(\Sigma_{-}) + \del_{x}(E_{1}^{1}N_{\times}) - N_{\times}\del_{x}(E_{1}^{1}) &= (q+3\Sigma_{+}-2)\Sigma_{-} + 2\sqrt{3}\Sigma_{\times}^{2} 
+ \sqrt{3}(\Sigma_{3}^{2} -\Sigma_{2}^{2}) \nonumber \\
&- 2\sqrt{3}N_{-}^{2} + \frac{\sqrt{3}}{2}(\tilde{T}^{22}-\tilde{T}^{33}), \\
\label{eqn:NTimes_ArealGauge}
\del_{t}(N_{\times}) + \del_{x}(E_{1}^{1}\Sigma_{-}) - \Sigma_{-}\del_{x}(E_{1}^{1}) &= (q+3\Sigma_{+})N_{\times}, \\
\label{eqn:SigmaTimes_ArealGauge}
\del_{t}(\Sigma_{\times}) - \del_{x}(E_{1}^{1}N_{-}) + N_{-}\del_{x}(E_{1}^{1}) &= (q+3\Sigma_{+}-2)\Sigma_{\times} -2\sqrt{3}N_{\times}N_{-} - 2\sqrt{3}\Sigma_{\times}\Sigma_{-} \nonumber \\
&+ 2\sqrt{3}\Sigma_{2}\Sigma_{3} + \sqrt{3}\tilde{T}^{23}, \\
\label{eqn:NMinus_ArealGauge}
\del_{t}(N_{-}) - \del_{x}(E_{1}^{1}\Sigma_{\times}) + \Sigma_{\times}\del_{x}(E_{1}^{1}) &= (q+3\Sigma_{+})N_{-} + 2\sqrt{3}\Sigma_{\times}N_{\times} + 2\sqrt{3}N_{-}\Sigma_{-}, \\
\label{eqn:SigmaPlus_ArealGauge}
\del_{t}(\Sigma_{+}) +\frac{1}{3}\del_{x}(E_{1}^{1}\dot{U}_{1}) - \dot{U}_{1}\del_{x}(E_{1}^{1}) &= -(q+3\Sigma_{+})(1-\Sigma_{+}) + 2(\Sigma_{+} +\Sigma_{-}^{2} + \Sigma_{\times}^{2}+\Sigma_{2}^{2}+\Sigma_{3}^{2}) \nonumber \\
&+\frac{1}{2}(\tilde{T}^{00}+\tilde{T}{_A^A}), \\
\label{eqn:Sigma2_ArealGauge}
\del_{t}(\Sigma_{2}) &= (q-2+\sqrt{3}\Sigma_{-})\Sigma_{2} -2\sqrt{3}\Sigma_{3}\Sigma_{\times} +\sqrt{3}\tilde{T}^{13},\\
\label{eqn:Sigma3_ArealGauge}
\del_{t}(\Sigma_{3}) &= (q-2-\sqrt{3}\Sigma_{-})\Sigma_{3}  + \sqrt{3}\tilde{T}^{12},\\
\label{eqn:E11_ArealGauge}
\del_{t}(E_{1}^{1}) &= (q+3\Sigma_{+})E_{1}^{1},
\end{align}
and the constraints reduce to
\begin{align}
\label{eqn:CM2_ArealGauge}
0 &= (\mathcal{C}_{M})_{2} := (E_{1}^{1}\del_{x}-r)\Sigma_{3} - \sqrt{3}\Sigma_{3}N_{\times} + \sqrt{3}\tilde{T}^{02}, \\
\label{eqn:CM3_ArealGauge}
0 &= (\mathcal{C}_{M})_{3} := (E_{1}^{1}\del_{x}-r)\Sigma_{2} + \sqrt{3}\Sigma_{2}N_{\times} + 2\sqrt{3}\Sigma_{3}N_{-} + \sqrt{3}\tilde{T}^{03}, \\
\label{eqn:CH_ArealGauge}
0 &= \mathcal{C}_{H} := 1 -2\Sigma_{+}-\Sigma_{\times}^{2}-\Sigma_{-}^{2}-\Sigma_{2}^{2}-\Sigma_{3}^{2}-N_{-}^{2}-N_{\times}^{2} -\tilde{T}^{00}.
\end{align}
Additionally, we have the following expressions for the auxiliary variables $q$ and $r$ 
\begin{align}
\label{eqn:q_expression_ArealGauge}
    q  &= \frac{1}{2} + \frac{3}{2}(\Sigma_{\times}^{2}+\Sigma_{-}^{2}+\Sigma_{2}^{2}+\Sigma_{3}^{2}) - 3(\Sigma_{2}^{2}+\Sigma_{3}^{2}) + \frac{3}{2}(N_{-}^{2}+N_{\times}^{2}) + \frac{3}{2}\tilde{T}^{11}, \\
\label{eqn:r_expression_ArealGauge}
    r &= 3\Sigma_{\times}N_{-}-3N_{\times}\Sigma_{-} -\frac{3}{2}\tilde{T}^{01}.
\end{align}
Next, recall the stress-energy tensor for a perfect fluid
\begin{align*}
    T^{ab} &= (\rho+p)v^{a}v^{b} + pg^{ab}.
\end{align*}
The 3+1 split of a perfect fluid, in terms of our frame, is given by
\begin{equation}
\begin{aligned}
\label{eqn:Tab_3+1_lorentz}
T^{00} &= (\Gamma^{2}(K+1)-K)\rho, \\
T^{0A} &= \Gamma^{2}(K+1)\rho\nu^{A} = (T^{00}+K\rho)\nu^{A} , \\
T^{AB} &= \Gamma^{2}(K+1)\rho\nu^{A}\nu^{B} + K\rho\eta^{AB} = (T^{00}+K\rho)\nu^{A}\nu^{B} + K\rho\eta^{AB}.
\end{aligned}
\end{equation}
The spatial fluid velocity $\nu_{A}$ and the Lorentz factor $\Gamma$ are defined as follows 
\begin{align}
\label{eqn:Fluid_3+1_decomp}
v^{a} = \Gamma(n^{a}+\nu^{a}), \quad |\nu|^{2} = \delta^{AB}\nu_{A}\nu_{B}, \quad \Gamma = \frac{1}{(1-|\nu|^{2})^{\frac{1}{2}}}, \quad |\nu| < 1,
\end{align}
where $n=e_{0}$ is the normal vector of the foliation and $n^{a}\nu_{a} = 0$.  The conservation of stress-energy 
\begin{align*}
    \nabla_{a}T^{ab} = 0.
\end{align*}
is then equivalent to the Euler equations
\begin{align}
\label{eqn:Euler_00_ArealGauge}
    \del_{t}(\tilde{T}^{00}) + \del_{x}(E_{1}^{1}\tilde{T}^{01}) - \tilde{T}^{01}\del_{x}(E_{1}^{1}) &= (2q -1 + 3\Sigma_{+})\tilde{T}^{00} +(3\Sigma_{+}-1)\tilde{T}^{11} \nonumber \\
    &- (1 +\sqrt{3}\Sigma_{-})\tilde{T}^{22} 
    - (1  - \sqrt{3}\Sigma_{-})\tilde{T}^{33} \nonumber \\
    &- 2\sqrt{3}\Sigma_{3}\tilde{T}^{12} - 2\sqrt{3}\Sigma_{2}\tilde{T}^{13} - 2\sqrt{3}\Sigma_{\times}\tilde{T}^{23}, \\
\label{eqn:Euler_01_ArealGauge}
    \del_{t}(\tilde{T}^{01}) + \del_{x}(E_{1}^{1}\tilde{T}^{11}) - \tilde{T}^{01}\del_{x}(E_{1}^{1}) &= 2(q -1 + 3\Sigma_{+})\tilde{T}^{01} + r(\tilde{T}^{11} - \tilde{T}^{00}) \nonumber \\
    &- 2\sqrt{3}(\Sigma_{3}\tilde{T}^{02} + \Sigma_{2}\tilde{T}^{03}) - \sqrt{3}N_{\times}(\tilde{T}^{22} - \tilde{T}^{33}) \nonumber \\
    &+ 2\sqrt{3}N_{-}\tilde{T}^{23}, \\
\label{eqn:Euler_02_ArealGauge}
    \del_{t}(\tilde{T}^{02}) + \del_{x}(E_{1}^{1}\tilde{T}^{12}) - \tilde{T}^{12}\del_{x}(E_{1}^{1}) &= (2q -2 + 3\Sigma_{+} - \sqrt{3}\Sigma_{-})\tilde{T}^{02} + (r + \sqrt{3}N_{\times} )\tilde{T}^{12}, \\
\label{eqn:Euler_03_ArealGauge}
    \del_{t}(\tilde{T}^{03}) + \del_{x}(E_{1}^{1}\tilde{T}^{13}) - \tilde{T}^{13}\del_{x}(E_{1}^{1}) &= (2q - 2 + 3\Sigma_{+} + \sqrt{3}\Sigma_{-})\tilde{T}^{03} + (r - \sqrt{3}N_{\times} )\tilde{T}^{13}  \nonumber \\
    &- 2\sqrt{3}\Sigma_{\times}\tilde{T}^{02} 
    - 2\sqrt{3}N_{-}\tilde{T}^{12}.
\end{align}
where $\tilde{T}^{ab}$ denotes the $\beta$-normalised frame components of the stress-energy tensor. Observe that we can either evolve \eqref{eqn:SigmaPlus_ArealGauge} or algebraically solve \eqref{eqn:CH_ArealGauge} to obtain $\Sigma_{+}$. We will always do the latter as this eliminates both an evolution equation and a constraint. \newline \par

It is natural to ask which spatially homogeneous Bianchi models are `included' in the $\mathbb{T}^{2}$-symmetric equations \eqref{eqn:SigmaMinus_ArealGauge}-\eqref{eqn:E11_ArealGauge}, \eqref{eqn:Euler_00_ArealGauge}-\eqref{eqn:Euler_03_ArealGauge}, as this will affect the spatially homogeneous dynamics which can be numerically observed. As discussed in\footnote{See also the extensive discussion in \cite{Lim_Thesis:2004}*{Chapter 3}.} \cite{EllisWainwright:1997}*{\S 12.4}, all Bianchi models which admit an Abelian $G_{2}$ subgroup are included in the general class of $G_{2}$ cosmologies\footnote{That is, cosmologies which admit a two-dimensional isometry group which acts on spacelike hypersurfaces.}. These models are given in Table 12.4 of \cite{EllisWainwright:1997}. However, since we use the timelike areal gauge, the Class B Bianchi cosmologies are \textit{not} included as a subset of the $\beta$-normalised equations presented here\footnote{This is due to the fact the frame variable $A_{A}=0$ in the timelike areal gauge.}. Thus, the Bianchi models included in the $\beta$-normalised equations in timelike areal gauge are types $I$, $II$, $VI_{0}$, and $VII_{0}$. As discussed in \cite{Lim_Thesis:2004}, the class B models could be included by working with the separable areal gauge instead of the timelike areal gauge, however we have not investigated this possibility as the frame equations are not obviously well-posed in the former gauge.

\subsection{Characteristic Speeds}
The gravitational variables form a symmetric hyperbolic system augmented with ODEs. It is straightforward to verify that the characteristic speeds are all given by $\pm E_{1}^{1}$. Let us now consider the Euler equations. After a short calculation, we see that the principal part of \eqref{eqn:Euler_00_ArealGauge}-\eqref{eqn:Euler_03_ArealGauge} can be expressed as 
\begin{align*}
    \del_{t}U + E_{1}^{1}\del_{x}F = \cdots
\end{align*}
where
\begin{align*}
    U &:= (\tilde{T}^{00},\tilde{T}^{01},\tilde{T}^{02},\tilde{T}^{03})^{\text{T}} \\
    F &:= (\tilde{T}^{01},\tilde{T}^{11},\tilde{T}^{12},\tilde{T}^{13})^{\text{T}}. 
\end{align*}
In order to compute the characteristic speeds, we will express the principal part as quasi-linear system in terms of the primitive variables $W := (\mu,\nu^{1},\nu^{2},\nu^{3})$, where $\mu$ is the $\beta$-normalised fluid density,
\begin{align*}
    A^{0}\del_{t}W + A^{1}\del_{x}W = \cdots
\end{align*}
where
\begin{align*}
    A^{0} &:= \frac{\del U}{\del W}, \\
    A^{1} &:= E_{1}^{1}\frac{\del F}{\del W} 
\end{align*}
The characteristic speeds are given by the eigenvalues of the flux Jacobian $\mathbf{F}$, 
\begin{align*}
    \mathbf{F} = E_{1}^{1}\frac{\del F}{\del U} = E^{1}_{1}\frac{\del F}{\del W}\frac{\del W}{\del U} = A^{1}(A^{0})^{-1}.
\end{align*}
After a short calculation, we see that the characteristic speeds are 
\begin{equation}
\begin{aligned}
\label{eqn:Fluid_CS}
    \lambda_{0} &= E^{1}_{1}\nu^{1}, \\
    \lambda_{\pm} &= E^{1}_{1}\frac{(1-K)\nu^{1} \pm \sqrt{K(1-|\nu|^{2})\Big(1-(\nu^{1})^{2} - K\big((\nu^{2})^{2} + (\nu^{3})^{2}\big)\Big)}}{1 - K|\nu|^{2}}.
\end{aligned}
\end{equation}
Note $\lambda_{0}$ has multiplicity two.

\subsection{Primitive Form of the Euler Equations}
\label{sec:Primitive_Euler}
In order to evolve the fluid equations \eqref{eqn:Euler_00_ArealGauge}-\eqref{eqn:Euler_03_ArealGauge}, it is necessary to recover the primitive variables $(\nu^{A},\mu)$ at each time step. In practice, since the normalised variables $\tilde{T}^{0a}$ rapidly decay to zero, the primitive recovery step becomes highly error prone. In particular, the normalisation constraint $|\nu|<1$ for the fluid velocity is often violated which leads to the code crashing. To resolve this, we directly evolve the primitive variables which allows us to easily enforce the normalisation constraint at each time step (see Section \ref{sec:Numerical_Setup} for details).  \newline \par

To derive evolution equations for the primitive variables from \eqref{eqn:Euler_00_ArealGauge}-\eqref{eqn:Euler_03_ArealGauge}, we first note that the following identities hold
\begin{align}
\label{eqn:T00_Ident_Prim_Derivation}
    \tilde{T}^{00} &= (K+1)\Gamma^{2}\mu - K\mu = \Gamma^{2}\mu(1+K|\nu|^{2}), \\
\label{eqn:T0A_Ident_Prim_Derivation}
    \tilde{T}^{0A} &= \frac{K+1}{1+K|\nu|^{2}}\tilde{T}^{00}\nu^{A}, \\
\label{eqn:TAB_Ident_Prim_Derivation}
    \tilde{T}^{AB} &= \frac{K+1}{1+K|\nu|^{2}}\tilde{T}^{00}\nu^{A}\nu^{B} + \frac{K\tilde{T}^{00}}{\Gamma^{2}(1+K|\nu|^{2})}\eta^{AB}.
\end{align}
Following \cites{Lim_Thesis:2004,Elst_et_al:2001}, we introduce the normalised derivative operators
\begin{align*}
    \Bdel_{0} = \del_{t}, \quad \Bdel_{1} = E_{1}^{1}\del_{x}.
\end{align*}
Expanding $\Bdel_{0}\big(\tilde{T}^{0A}\tensor{\tilde{T}}{^0_A}\big)$ using \eqref{eqn:T0A_Ident_Prim_Derivation}, we find
\begin{align*}
    (K+1)^{2}(\tilde{T}^{00})^{2}\frac{1-K|\nu|^{2}}{(1+K|\nu|^{2})^{3}}\Bdel_{0}(|\nu|^{2}) = \Bdel_{0}(\tilde{T}^{0A}\tensor{\tilde{T}}{^0_A}) - 2\tilde{T}^{00}(K+1)^{2}\frac{|\nu|^{2}}{(1+K|\nu|^{2})^{2}}\Bdel_{0}(\tilde{T}^{00}),
\end{align*}
which can be re-arranged to obtain
\begin{align}
\label{eqn:nu_norm_evo}
    \Bdel_{0}(|\nu|^{2}) = \frac{(1+K|\nu|^{2})}{(K+1)\tilde{T}^{00}(1-K|\nu|^{2})}\Big(2\nu_{A}\Bdel_{0}(\tilde{T}^{0A}) - \frac{2(K+1)|\nu|^{2}}{1+K|\nu|^{2}}\Bdel_{0}(\tilde{T}^{00})\Big).
\end{align}
Now, using \eqref{eqn:T0A_Ident_Prim_Derivation}, we find that
\begin{align}
\label{eqn:nuA_evo1}
    \Bdel_{0}(\nu^{A}) &= \frac{(1+K|\nu|^{2})}{(K+1)\tilde{T}^{00}}\Bdel_{0}(\tilde{T}^{0A}) -\frac{(1+K|\nu|^{2})\tilde{T}^{0A}}{(K+1)(\tilde{T}^{00})^{2}}\Bdel_{0}(\tilde{T}^{00}) + \frac{K\tilde{T}^{0A}}{(K+1)\tilde{T}^{00}}\Bdel_{0}(|\nu|^{2}) \nonumber \\
    &= \frac{(1+K|\nu|^{2})}{(K+1)\tilde{T}^{00}}\Bdel_{0}(\tilde{T}^{0A}) -\frac{\nu^{A}}{\tilde{T}^{00}}\Bdel_{0}(\tilde{T}^{00}) + \frac{K\nu^{A}}{1+K|\nu|^{2}}\Bdel_{0}(|\nu|^{2}).
\end{align}
Substituting \eqref{eqn:nu_norm_evo} into \eqref{eqn:nuA_evo1} then yields
\begin{align}
\label{eqn:nuA_evo2}
    \Bdel_{0}(\nu^{A}) &= \frac{1+K|\nu|^{2}}{(K+1)\tilde{T}^{00}(1-K|\nu|^{2})}\Bigg[\Big((1-K|\nu|^{2})\delta^{A}_{B} + 2K\nu^{A}\nu_{B}\Big)\Bdel_{0}(\tilde{T}^{0B}) \nonumber \\
    &- (K+1)\nu^{A}\Bdel_{0}(\tilde{T}^{00})\Bigg].
\end{align}
Now, by systematically replacing the stress-energy terms in \eqref{eqn:Euler_00_ArealGauge} and \eqref{eqn:nuA_evo2} using \eqref{eqn:Euler_00_ArealGauge}-\eqref{eqn:Euler_03_ArealGauge} and \eqref{eqn:T00_Ident_Prim_Derivation}-\eqref{eqn:TAB_Ident_Prim_Derivation}, we obtain the following system of equations
\begin{align}
\label{eqn:log_T00_evo}
    \Bdel_{0}\big(\log(\tilde{T}^{00})\big) &= -\frac{(1+K)\nu^{1}}{1+K|\nu|^{2}}\Bdel_{1}\big(\log(\tilde{T}^{00})\big) - \frac{(1+K)(1+K|\nu|^{2} -2K(\nu^{1})^{2})}{(1+K|\nu|^{2})^{2}}\Bdel_{1}(\nu^{1}) \nonumber \\ 
    &+ \frac{2K(1+K)\nu^{1}\nu^{2}}{(1+K|\nu|^{2})^{2}}\Bdel_{1}(\nu^{2}) + \frac{2K(1+K)\nu^{1}\nu^{3}}{(1+K|\nu|^{2})^{2}}\Bdel_{1}(\nu^{3}) \nonumber \\
    &+ \frac{1}{1+K|\nu|^{2}}\Big(-1 -3K + 2q + 2K(1+q)|\nu|^{2} 
    + (K+1)(-1 + \sqrt{3}\Sigma_{-})(\nu^{3})^{2}  
    \nonumber \\
    &- (1+K)(1+\sqrt{3}\Sigma_{-})(\nu^{2})^{2} + 3(K+1)\Sigma_{+} + (1+K)(-1+\sqrt{3}\Sigma_{+})(\nu^{1})^{2} \nonumber \\
    &- 2\sqrt{3}(1+K)\nu^{1}(\nu^{3}\Sigma_{2} + \nu^{2}\Sigma_{3}) -2\sqrt{3}(1+K)\nu^{2}\nu^{3}\Sigma_{\times}  \Big), \\
\label{eqn:nu1_evo}
    \Bdel_{0}(\nu^{1}) &= \frac{K}{(1+K)(-1+K|\nu|^{2})}\Big(1+K|\nu|^{4} + (\nu^{1})^{2}\big(-1 +K +2(1+K)|\nu|^{2}\big) \nonumber \\
    &- |\nu|^{2}\big(1+K + (1+3K)(\nu^{1})^{2}\big)\Big)\Bdel_{1}\big(\log(\tilde{T}^{00})\big) \nonumber \\
    &+ \frac{\nu^{1}(1 - 3K + (1+K)K|\nu|^{2} + 2K(1-K)(\nu^{1})^{2})}{-1+K^{2}|\nu|^{4}}\Bdel_{1}(\nu^{1}) \nonumber \\
    &+ \frac{2K\big(-1+K|\nu|^{2} - (K-1)(\nu^{1})^{2}\big)\nu^{2}}{-1 + K^{2}|\nu|^{4}}\Bdel_{1}(\nu^{2}) \nonumber \\
    &+ \frac{2K\big(-1+K|\nu|^{2} - (K-1)(\nu^{1})^{2}\big)\nu^{3}}{-1 + K^{2}|\nu|^{4}}\Bdel_{1}(\nu^{3}) \nonumber \\
    &+ S_{1}, \\
\label{eqn:nu2_evo}
    \Bdel_{0}(\nu^{2}) &= \frac{K\nu^{1}\nu^{2}(1-K)(|\nu|^{2}-1)}{(1+K)(-1+K|\nu|^{2})}\Bdel_{1}\big(\log(\tilde{T}^{00})\big) \nonumber \\
    &+ \frac{K\nu^{2}\big((|\nu|^{2}-1)(1+K|\nu|^{2}) - 2(K-1)(\nu^{1})^{2}\big)}{-1+K^{2}|\nu|^{4}}\Bdel_{1}(\nu^{1}) \nonumber \\
    &+ \nu^{1}\Big(-1 - \frac{2(K-1)K(\nu^{2})^{2}}{-1+K^{2}|\nu|^{4}}\Big)\Bdel_{1}(\nu^{2}) + \frac{2(1-K)K\nu^{1}\nu^{2}\nu^{3}}{-1+K^{2}|\nu|^{4}}\Bdel_{1}(\nu^{3}) + S_{2}, \\
\label{eqn:nu3_evo}
    \Bdel_{0}(\nu^{3}) &= \frac{K\nu^{1}\nu^{3}(1-K)(|\nu|^{2}-1)}{(1+K)(-1+K|\nu|^{2})}\Bdel_{1}\big(\log(\tilde{T}^{00})\big) \nonumber \\
    &+ \frac{K\nu^{3}\big((|\nu|^{2}-1)(1+K|\nu|^{2}) - 2(K-1)(\nu^{1})^{2}\big)}{-1+K^{2}|\nu|^{4}}\Bdel_{1}(\nu^{1}) \nonumber \\
    & +\frac{2(1-K)K\nu^{1}\nu^{2}\nu^{3}}{-1+K^{2}|\nu|^{4}}\Bdel_{1}(\nu^{2}) + \nu^{1}\Big(-1 - \frac{2(K-1)K(\nu^{3})^{2}}{-1+ K^{2}|\nu|^{4}}\Big)\Bdel_{1}(\nu^{3}) + S_{3},
\end{align}
where the source terms $S_{1}$, $S_{2}$, and $S_{3}$ are given by
\begin{align*}
    S_{1} &= 2\sqrt{3}N_{-}\nu^{2}\nu^{3} + \sqrt{3}N_{\times}\big(-|\nu|^{2} + (\nu^{1})^{2} +2(\nu^{3})^{2}\big) + \frac{1}{(1+K)(-1+K|\nu|^{2})}\Bigg((1-K)r \nonumber \\ &- 2K^{2}r|\nu|^{4} + |\nu|^{2}\Big(K(1+K)r + (1+K)(3K-1)\nu^{1} + K(1+5K)r(\nu^{1})^{2}\Big) \nonumber \\
    &+ \nu^{1}\Big((1+K)(1-3K) + r\nu^{1}\big(-1 + K - 2K^{2} - 2K(1+K)|\nu|^{2}\big)\Big)\Bigg) \nonumber \\
    &+ \frac{\sqrt{3}(K-1)\nu^{1}\big(|\nu|^{2}-(\nu^{1})^{2}-2(\nu^{3})^{2}\big)\Sigma_{-}}{-1+K|\nu|^{2}} - \frac{3(K-1)\nu^{1}\big((\nu^{1})^{2}-1\big)\Sigma_{+}}{-1+K|\nu|^{2}} \nonumber \\ 
    &- \frac{2\sqrt{3}\big(-1+K|\nu|^{2} - (K-1)(\nu^{1})^{2}\big)\nu^{3}\Sigma_{2}}{-1+K|\nu|^{2}} - \frac{2\sqrt{3}\big(-1+K|\nu|^{2} - (K-1)(\nu^{1})^{2}\big)\nu^{2}\Sigma_{3}}{-1+K|\nu|^{2}} \nonumber \\
    &+ \frac{2\sqrt{3}(K-1)\nu^{1}\nu^{2}\nu^{3}\Sigma_{\times}}{-1+K|\nu|^{2}}, \\
    S_{2} &= \sqrt{3}N_{\times}\nu^{1}\nu^{2} + \frac{\big((1+K)(3K-1)(|\nu|^{2}-1) + r(-1 + K -2K^{2} + K(3K-1)|\nu|^{2})\nu^{1}\big)\nu^{2}}{(1+K)(-1+K|\nu|^{2})} \nonumber \\
    &+ \frac{1}{-1+K|\nu|^{2}}\Bigg(\sqrt{3}\nu^{2}\big(1-|\nu|^{2}-(K-1)(\nu^{1})^{2} - 2(K-1)(\nu^{3})^{2}\big)\Sigma_{-} \nonumber \\
    &+ 3\Big(K + (\nu^{1})^{2}-K\big(|\nu|^{2}+(\nu^{1})^{2}\big)\Big)\nu^{2}\Sigma_{+} + 2\sqrt{3}(K-1)\nu^{1}\nu^{2}\nu^{3}\Sigma_{2} \nonumber \\
    &+ 2\sqrt{3}(K-1)\nu^{1}(\nu^{2})^{2}\Sigma_{3} + 2\sqrt{3}(K-1)(\nu^{2})^{2}\nu^{3}\Sigma_{\times}\Bigg), \\
    S_{3} &= -2\sqrt{3}N_{-}\nu^{1}\nu^{2} - \sqrt{3}N_{\times}\nu^{1}\nu^{3} + \frac{1}{(1+K)(-1+K|\nu|^{2})}\Big((1+K)(3K-1)(|\nu|^{2}-1) \nonumber \\
    &+ r(-1 + K -2K^{2} + K(3K-1)|\nu|^{2})\nu^{1}\Big)\nu^{3} + \frac{1}{-1+K|\nu|^{2}}\Bigg(- \sqrt{3}\nu^{3}\big(1 + (1-2K)|\nu|^{2} \nonumber  \\
    &+ (K-1)(\nu^{1})^{2} + 2(K-1)(\nu^{3})^{2}\big)\Sigma_{-} + 3\Big(K+(\nu^{1})^{2}-K\big(|\nu|^{2}-(\nu^{1})^{2}\big)\Big)\nu^{3}\Sigma_{\times} \nonumber \\
    &+ 2\sqrt{3}(K-1)\nu^{1}(\nu^{3})^{2}\Sigma_{2} + 2\sqrt{3}(K-1)\nu^{1}\nu^{2}\nu^{3}\Sigma_{3} - 2\sqrt{3}\nu^{2}\big(-1+K|\nu|^{2}-(K-1)(\nu^{3})^{2}\big)\Sigma_{\times} \Bigg).
\end{align*}

\begin{rem}
    We could, of course, evolve $\log(\mu)$ instead of $\log(\tilde{T}^{00})$. However, since it is straightforward to obtain $\log(\mu)$ from $\log(\tilde{T}^{00})$ and $\tilde{T}^{00}$ appears in several other equations, we have not tested this alternative.
\end{rem}

To summarise, we numerically solve \eqref{eqn:SigmaMinus_ArealGauge}-\eqref{eqn:NMinus_ArealGauge}, \eqref{eqn:Sigma2_ArealGauge}-\eqref{eqn:E11_ArealGauge}, and \eqref{eqn:log_T00_evo}-\eqref{eqn:nu3_evo} subject to the constraints \eqref{eqn:CM2_ArealGauge}-\eqref{eqn:CM3_ArealGauge}. The auxiliary variables $\Sigma_{+}$, $q$, and $r$, are obtained from the expressions \eqref{eqn:CH_ArealGauge}, \eqref{eqn:q_expression_ArealGauge}, and \eqref{eqn:r_expression_ArealGauge} respectively.

\section{Spatially Homogeneous Asymptotics}
\label{sec:Homogeneous_Asymptotics}
In order to understand our numerical results, it is useful to first discuss the expected asymptotic behaviour near the singularity\footnote{See \cite{UEWE:2003} for a detailed discussion of the BKL conjecture and the generalised mixmaster attractor.}. Initially, the spatial derivatives, frame component $E_{1}^{1}$, and stress-energy components $\tilde{T}^{ab}$ are all non-negligible and the solution evolves in an inhomogeneous manner. This is followed by a regime in which $E_{1}^{1}$ and $\tilde{T}^{ab}$ exponentially decay towards zero. As these terms become suitably small, the gravitational dynamics at each point are well described by the spatially homogeneous equations obtained by setting the spatial derivatives and stress-energy tensor components to zero in \eqref{eqn:SigmaMinus_ArealGauge}-\eqref{eqn:NMinus_ArealGauge} and \eqref{eqn:Sigma2_ArealGauge}-\eqref{eqn:E11_ArealGauge}. Note that while the local (pointwise) dynamics are consistent with a spatially homogeneous cosmology, the overall spacetime remains inhomogeneous. In particular, the spatial derivative of any given variable is not necessarily small. However, these terms become negligible in the evolution equations because the spatial derivatives are all multiplied by the frame component $E_{1}^{1}$, which exponentially decays to zero.   \newline \par

A key part of the BKL conjecture is that the approach to the singularity is described (locally) by oscillatory mixmaster dynamics. This consists of \textit{epochs}, where the gravitational variables closely approximate a vacuum Kasner solution, and \textit{bounces}, where the solution rapidly transitions to a new Kasner solution with different Kasner exponents starting a new epoch. As mentioned in the introduction, incorporating a fluid introduces a new dynamical behaviour in the form of tilt transitions, where the solution can oscillate between orthogonal ($|\nu|=0$) and extremely tilted ($|\nu|=1$) Kasner states. It is important to emphasise that while gravitational dynamics can influence tilt transitions in the fluid, the reverse is not true. This is because the stress-energy tensor is negligible during the mixmaster regime and the evolution of the gravitational variables is essentially decoupled from the fluid. In particular, the fluid will behave like a test field on an oscillating Kasner background. As the gravitational dynamics have been discussed extensively in the literature, see for example \cites{UEWE:2003,Garfinkle:2004,Andersson_et_al:2005,Ringstrom:2001}, we will focus on the fluid dynamics for the remainder of this section. \newline \par 

In terms of our tetrad variables,  the Kasner solutions are described by
\begin{gather*}
N_{+} = N_{\times} = \Sigma_{2} = \Sigma_{3} = \Sigma_{\times} = \mu = 0, \\
\Sigma_{-} = \mathring{\Sigma}_{-}, \quad \Sigma_{+} = \mathring{\Sigma}_{+}, \\ 
|\nu| = 0 \;\; (\text{Orthogonal}) \quad \text{or} \quad |\nu| = 1 \;\; (\text{Extremely Tilted})
\end{gather*}
where $\mathring{\Sigma}_{-}$ and $\mathring{\Sigma}_{+}$ are constants satisfying
\begin{align}
\label{eqn:Kasner_Parabola}
    2\mathring{\Sigma}_{+} + \mathring{\Sigma}_{-}^{2} = 1
\end{align}
Note that \eqref{eqn:Kasner_Parabola} corresponds to the familiar `Kasner circle'
\begin{align*}
    (\mathring{\Sigma}_{+}^{H})^{2} + (\mathring{\Sigma}_{-}^{H})^{2} = 1,
\end{align*}
in terms of the Hubble-normalised variables $(\Sigma_{+}^{H},\Sigma_{-}^{H})$ 
\begin{align*}
    \Sigma_{+}^{H} = \frac{1}{(1-\Sigma_{+})}\Sigma_{+}, \quad \Sigma_{-}^{H} = \frac{1}{(1-\Sigma_{+})}\Sigma_{-}.
\end{align*}
Similarly, the Kasner exponents are given by \cite{EllisWainwright:1997}*{\S 6.2.2}
\begin{align*}
    P_{1} &:= \frac{1}{3}(1 - 2\Sigma^{H}_{+}), \\
    P_{2} &:= \frac{1}{3}(1 + \Sigma^{H}_{+} + \sqrt{3}\Sigma_{-}^{H}), \\
    P_{3} &:= \frac{1}{3}(1 + \Sigma^{H}_{+} - \sqrt{3}\Sigma_{-}^{H}).
\end{align*}

What causes a tilt transition in the fluid velocity? To determine this, we linearise the spatially homogeneous fluid equations about the orthogonal and extremely tilted Kasner solutions. First, we consider the spatially homogeneous equations for the spatial velocity $\nu^{A}$ obtained by setting the spatial derivatives in \eqref{eqn:nu1_evo}-\eqref{eqn:nu3_evo} to zero, 
\begin{align}
\label{eqn:nu1_spatially_homogeneous}
    \del_{t}(\nu^{1}) &= S_{1}, \\ 
\label{eqn:nu2_spatially_homogeneous}
    \del_{t}(\nu^{2}) &= S_{2}, \\ 
\label{eqn:nu3_spatially_homogeneous}
    \del_{t}(\nu^{3}) &= S_{3}.
\end{align}
Now, by linearising \eqref{eqn:nu1_spatially_homogeneous}-\eqref{eqn:nu3_spatially_homogeneous} about the orthogonal ($|\nu|=0)$ Kasner solution we obtain the equations
\begin{align*}
    \del_{t}\delta\nu^{1} &= \frac{3\delta\nu^{1}(K-P_{1})}{1+\mathring{\Sigma}_{+}^{H}}, \\
    \del_{t}\delta\nu^{2} &= \frac{3\delta\nu^{2}(K-P_{2})}{1+\mathring{\Sigma}_{+}^{H}}, \\
    \del_{t}\delta\nu^{3} &= \frac{3\delta\nu^{3}(K-P_{3})}{1+\mathring{\Sigma}_{+}^{H}}.
\end{align*}
Using these equations, we observe that the component $\delta\nu^{A}$ will grow towards the past\footnote{Recall that the singularity is located at $t=-\infty$.} if $K-P_{A} < 0$. Based on this heuristic analysis, if the solution is close to an orthogonal Kasner state satisfying, for example, $K-P_{1} < 0$, then we expect the corresponding spatial velocity component $\nu^{1}$ will begin to approach an extremely tilted state $\nu^{1} = \pm 1$. Linearising \eqref{eqn:nu1_spatially_homogeneous}-\eqref{eqn:nu3_spatially_homogeneous} about the extremely tilted Kasner solutions with
\begin{align*}
    \nu^{A} = \pm 1, \quad \nu^{B} = 0, \quad A \neq B,
\end{align*}
for $A \in \{1,2,3\}$ then leads to remaining conditions in Table \ref{tab:velocity_triggers}. We note that these agree with the triggers obtained by Uggla et al.\ \cite{UEWE:2003}. Similar conditions on the Kasner exponents have also been obtained in a recent non-linear stability result by Beyer and Oliynyk \cite{BeyerOliynyk:2024} (see also \cite{BeyerOliynyk:2024b}). As with the previous example, when a trigger condition is met, we expect the corresponding component of the fluid velocity will begin to evolve away from its current tilt state to the opposite tilt state. Notably, since all the trigger conditions involve the Kasner exponents, this suggests that oscillations in the gravitational variables will induce tilt transitions in the fluid velocity. 

\begin{table}
    \centering
    \begin{tabular}{|c|c|}
    \hline
Equilibrium Point & Trigger for $\nu^{A}$ \\ \hline
         Kasner Circle ($|\nu|=0$) & $P_{A} - K > 0$ \\ \hline
         Extreme Tilt Kasner ($\nu^{A} = \pm 1$) & $P_{A} - K < 0$ \\ \hline
         Extreme Tilt Kasner ($\nu^{B} = \pm 1$, $A\neq B$) &  $P_{A} > P_{B}$  \\ \hline
    \end{tabular}
    \caption{Triggers for $\nu^{A}$ tilt transitions on the orthogonal and extreme tilt Kasner circles. $P_{A}$ is the corresponding Kasner exponent. $A,B \in \{1,2,3\}$. }
    \label{tab:velocity_triggers}
\end{table}

\section{Numerical Setup}
\label{sec:Numerical_Setup}
We now discuss our numerical scheme for solving the Einstein-Euler equations. The spatial computational domain is taken to be $[0,2\pi]$ which is discretised into a uniform grid with $N$ cells $C_{j} = [x_{j-\frac{1}{2}},x_{j+\frac{1}{2}}]$ and periodic boundary conditions.

\begin{rem}
    While periodic boundary conditions are standard in numerical simulations of cosmological models, the assumption of periodicity necessarily restricts the choice of initial data \cite{Aurrekoetxea_et_al:2023}. Moreover, several recent works have shown that enforcing periodicity can influence the evolution of inhomogeneous cosmological models \cites{Garfinkle_et_al:2023,ColeyLim:2023,Racz_et_al:2021}. In particular, recent work by Coley and Lim \cite{ColeyLim:2023}, has demonstrated that periodic initial data suppresses the growth of spatial curvature. This suggests that the boundary conditions could play a non-trivial role during the inhomogeneous stage of the evolution which, in turn, may influence the state of the solution when mixmaster dynamics begin\footnote{We emphasise that the boundary conditions should only affect the solution during the initial inhomogeneous stage of the evolution because the mixmaster regime is local (i.e. ODE dominated).}. One alternative is to implement an excision-type boundary, such as the `zooming' technique used in \cites{ColeyLim:2023, Lim:2009}, however we have not investigated this in the present article.
\end{rem}

Since we expect discontinuities to form in the fluid, it is necessary to implement some form of shock-capturing scheme. Typically, these schemes rely on expressing the equations in a flux-conservative form
\begin{align*}
    \del_{t}U + \del_{x}F = S,
\end{align*}
where $U$ are the conserved variables, $F$ is the flux, and $S$ is a source term. Unfortunately the Einstein-Euler equations cannot be expressed in this form with respect to our orthonormal frame. This is due to the frame component $E_{1}^{1}$ which appears in front of the spatial derivatives. For example, the evolution equations \eqref{eqn:Euler_00_ArealGauge}-\eqref{eqn:Euler_03_ArealGauge} for the frame analogues of the standard conserved variables $\tilde{U} = (\tilde{T}^{00},\tilde{T}^{0A})$ can be written schematically as
\begin{align*}
    \del_{t}\tilde{U} + E_{1}^{1}\del_{x}F = S.
\end{align*}
In particular, the flux term $E_{1}^{1}\del_{x}F$ cannot be written as a total divergence. Our initial approach was to split this term into a total divergence and the spatial derivative of the frame as follows 
\begin{align}
\label{eqn:pseudo_FC_eqn}
    \del_{t}\tilde{U} + \del_{x}(E_{1}^{1}F) - F\del_{x}(E_{1}^{1}) = S.
\end{align}
A local Lax-Friedrichs discretisation was used for the divergence terms while finite difference stencils were used to discretise the spatial derivative of the frame. While this approach does not have the same mathematical justification as conservative schemes, it did allow us to stably evolve discontinuities in the fluid. \newline \par

A more fundamental problem with this approach, however, is that the variables $(\tilde{T}^{00},\tilde{T}^{0A})$ rapidly decay to zero as the singularity is approached. In particular, as the `conserved' variables become small, recovering the primitive variables\footnote{It is necessary to recover the primitive variables at each time step in order to evaluate the flux and source terms.} $(\mu,\nu^{A})$ becomes highly error prone. This, in turn, consistently leads to violations of the velocity constraint 
\begin{align}
\label{eqn:spatial_fluid_constraint}
    |\nu|<1,
\end{align} 
which must be preserved to maintain the well-posedness of the fluid equations and ensure the solution is physical. Indeed, all numerical schemes we have tested inevitably crash if $|\nu| \geq 1$. \newline \par

As mentioned in Section \ref{sec:Primitive_Euler}, we avoid the issues with primitive recovery by directly evolving the variables  $(\log(\tilde{T}^{00}),\nu^{A})$ using \eqref{eqn:log_T00_evo}-\eqref{eqn:nu3_evo}. However, as before, the evolution equations for these variables cannot be written in flux-conservative form. Moreover, it does not appear possible to express them in the form \eqref{eqn:pseudo_FC_eqn}, where we could combine a standard shock-capturing scheme with finite differences for the frame derivatives, as described above. Our solution is to numerically evolve \eqref{eqn:log_T00_evo}-\eqref{eqn:nu3_evo} using a \textit{path-conservative} finite volume scheme\footnote{See \cite{Pares:2006} for a review of path-conservative schemes.}. These schemes, based on the theory of weak solutions for non-conservative hyperbolic systems developed by Dal Maso, LeFloch, and Murat (DLM) \cite{DLM:1995}, provide a natural way to stably evolve discontinuities in the fluid without working in a conservative form. 

\subsection{Path-Conservative Scheme}
One of the key challenges with non-conservative hyperbolic systems is that the usual definition of a weak solution in terms of distributions cannot be used. For a conservation law of the form,
\begin{align*}
    \del_{t}U + \del_{x}(F) = 0,
\end{align*}
a weak solution must satisfy the Rankine-Hugoniot conditions across a discontinuity
\begin{align}
\label{eqn:RankineHugoniot_Conservative}
    s(u_{r} - u_{l}) = \int^{x_{r}}_{x_{l}}\del_{x}\big(F(u)\big) \; dx = F(u_{r}) - F(u_{l}),
\end{align}
where $s$ is the shock speed. However, for a non-conservative system such as 
\begin{align*}
    \del_{t}u + A(u)\del_{x}u = 0
\end{align*}
the non-conservative product $A(u)\del_{x}u$ is not well-defined in the sense of the distributions. In the theory\footnote{See \cite{DLM:1995} for a mathematically rigorous presentation.} of DLM, given a Lipschitz continuous path $\Psi$ connecting the states $u_{r}$ and $u_{l}$
\begin{align}
\label{eqn:path_function_defn}
   \Psi: [0,1] \times \mathbb{R}^{N} \times \mathbb{R}^{N} \rightarrow \mathbb{R}^{N}, \quad  \Psi(0;u_{l},u_{r}) = u_{l}, \quad \Psi(1;u_{l},u_{r}) = u_{r},
\end{align}
the non-conservative product $A(u)\del_{x}u$ is defined as a Borel measure. This, in turn, allows one to rigorously define a notion of weak solutions for non-conservative hyperbolic systems. In particular, we obtain the generalised Rankine-Hugoniot conditions
\begin{align}
\label{eqn:RankineHugoniot_PathConservative}
    s(u_{r} - u_{l}) = \int^{1}_{0}A\big(\Psi(s;u_{l},u_{r})\big)\del_{s}\Psi(s;u_{l},u_{r})\; ds.
\end{align}
Observe that we recover the standard Rankine-Hugoniot conditions \eqref{eqn:RankineHugoniot_Conservative} in the case $A(u)$ is the Jacobian of a flux function $f(u)$. \newline \par

It is important to emphasise that this generalised notion of a weak solution depends on the family of paths used. Determining an appropriate choice of path is, in general, a difficult problem which we will not address here. Moreover, it is not guaranteed that a path-conservative scheme will converge to a weak solution corresponding to the given path\footnote{However, see \cite{Castro_et_al:2013} for an implementation of path-conservative schemes which discretely satisfy entropy conditions.} \cites{AbgrallKarni:2010,Castro_et_al:2008}. This may appear problematic, however this is not particularly different from numerical schemes for conservative systems, which  may also not converge to the physically relevant weak solution\footnote{Worse still, for systems of conservation laws in two or three space dimensions, even entropy solutions may not be unique! See for example \cites{Chiodaroli_et_al:2014,Chiodaroli:2014}.}. In any case, since we are primarily interested in the mixmaster regime, for which for the dynamics are ODE-dominated, the path-conservative aspect of our scheme will only play a small role in the initial inhomogeneous part of the evolution.  \newline \par

For simplicity, we employ a central-upwind path-conservative scheme inspired by \cite{DiazKurganovdeLuna:2019} to evolve both the gravitational and fluid variables. Schematically, the evolution equations \eqref{eqn:SigmaMinus_ArealGauge}-\eqref{eqn:NMinus_ArealGauge}, \eqref{eqn:Sigma2_ArealGauge}-\eqref{eqn:E11_ArealGauge}, and \eqref{eqn:log_T00_evo}-\eqref{eqn:nu3_evo} can be written as\footnote{Note that for the fluid equations \eqref{eqn:log_T00_evo}-\eqref{eqn:nu3_evo}, the flux term $F(w)=0$.}
\begin{align}
\label{eqn:Einstein_Euler_SchematicForm}
    \del_{t}w + \del_{x}F(w) - B(w)\del_{x}w = S,
\end{align}
where $w$, $F(w)$, and $S$ are vectors with $M$ components and $B(w)$ is an $M\times M$ matrix. Our numerical scheme for solving \eqref{eqn:Einstein_Euler_SchematicForm} can be expressed in the following semi-discrete form
\begin{align}
\label{eqn:PC_Scheme}
  \frac{d}{dt}\bar{w}_{j} &= \frac{-1}{\Delta x}\Big[ H_{j+\frac{1}{2}} - H_{j-\frac{1}{2}} - B_{j} - \frac{1}{2}\big(B_{\Psi,j+\frac{1}{2}} + B_{\Psi,j-\frac{1}{2}}\big)\Big] + S_{j}, 
\end{align}
where\footnote{Observe that the sign of the dissipation term $ a_{j+\frac{1}{2}}(w^{+}_{j+\frac{1}{2}} - w^{-}_{j+\frac{1}{2}})$ in the numerical flux \eqref{eqn:numerical_flux} is positive since we are evolving towards the past.} 
\begin{align}
\label{eqn:numerical_flux}
    H_{j+\frac{1}{2}} &:= \frac{1}{2}\big[ F(w^{+}_{j+\frac{1}{2}}) + F(w^{-}_{j+\frac{1}{2}}) + a_{j+\frac{1}{2}}(w^{+}_{j+\frac{1}{2}} - w^{-}_{j+\frac{1}{2}})\big],\\
    \label{eqn:B_j_defn}
    B_{j} &:= \int_{C_{j}}B\big(P_{j}(x)\big) \Big(\frac{d P^{(1)}_{j}}{dx},...,\frac{d P^{(M)}_{j}}{dx}\Big)^{\text{T}} \; dx, \\
\label{eqn:B_Psi_defn}
    B_{\Psi,j+\frac{1}{2}} &:= \int^{1}_{0}B\big(\Psi_{j-\frac{1}{2}}(s)\big)\Big(\frac{d\Psi^{(1)}_{j-\frac{1}{2}}}{ds}, ... ,\frac{d\Psi^{(M)}_{j-\frac{1}{2}}}{ds}\Big)^{\text{T}} \; ds , \\
S_{j} &:= S(\bar{w}_{j}), 
\end{align}
and, following \cite{DiazKurganovdeLuna:2019}, we use the simple segment path 
\begin{align*}
    \Psi(s)_{j+\frac{1}{2}} = w^{-}_{j+\frac{1}{2}} + s\big(w^{+}_{j+\frac{1}{2}} - w^{-}_{j+\frac{1}{2}}\big).
\end{align*}
In the above, $\bar{w}_{j}$ denotes the vector of cell-averages, $w^{-}_{j+\frac{1}{2}}$ and $w^{+}_{j+\frac{1}{2}}$ are the piecewise linear reconstructions approaching from the left and the right respectively,
\begin{align*}
    w^{-}_{j+\frac{1}{2}} &= P_{j}(x_{j+\frac{1}{2}}), \\
    w^{+}_{j+\frac{1}{2}} &= P_{j+1}(x_{j+\frac{1}{2}}), \\
    P_{j}(x) &= \bar{w}_{j} + (w_{x})_{j}(x-x_{j}),
\end{align*}
$(w_{x})_{j+\frac{1}{2}}$ is a minmod limited approximation of the derivative 
\begin{align*}
    (w_{x})_{j} &= \frac{1}{\Delta x}\phi(r_{j})(\bar{w}_{j+1} -\bar{w}_{j}), \\
    r_{j} &= \frac{\bar{w}_{j} - \bar{w}_{j-1}}{\bar{w}_{j+1} - \bar{w}_{j}}, \\
    \phi(r) &= \max\big(0,\min(r,1)\big),
\end{align*}
and $a_{j+\frac{1}{2}}$ is the maximum local characteristic speed\footnote{Note that $\lambda^{\pm}_{j+\frac{1}{2}}$ denote the eigenvalues evaluated using the reconstructions $w^{\pm}_{j+\frac{1}{2}}$ not the fluid characteristic speeds $\lambda_{\pm}$ defined in \eqref{eqn:Fluid_CS}.}
\begin{align*}
    a_{j+\frac{1}{2}} = \max\big(|\lambda^{-}_{j+\frac{1}{2}}|,|\lambda^{+}_{j+\frac{1}{2}}|\big).
\end{align*}
By using the midpoint rule to evaluate the integrals \eqref{eqn:B_j_defn}-\eqref{eqn:B_Psi_defn}, the terms $B_{j}$ and $B_{\Psi,j+\frac{1}{2}}$ become
\begin{align*}
    B_{j} &= B(\bar{w}_{j})(w^{-}_{j+\frac{1}{2}} - w^{+}_{j-\frac{1}{2}}), \\
    B_{\Psi,j+\frac{1}{2}} &= B(\hat{w}_{j})(w^{+}_{j+\frac{1}{2}} - w^{-}_{j+\frac{1}{2}}), \\
    \hat{w}_{j} &= \frac{1}{2}(w^{-}_{j+\frac{1}{2}} + w^{+}_{j+\frac{1}{2}}), 
\end{align*}
The resulting scheme is formally second-order accurate on smooth solutions\footnote{Near discontinuities, the solution is expected to be approximately first-order accurate however it may be worse, see \cite{LeVeque:2002}*{\S 8.7} for details.}. We integrate in time using a fourth-order Runge-Kutta method. The time step is taken to be
\begin{align*}
    \Delta t = \min\{C\Delta x, C\frac{\Delta x}{|\lambda_{\max}|}\}
\end{align*}
where $C$ is the CFL constant, typically taken to be $0.01$, and  $|\lambda_{\max}|$ is the largest value of the characteristic speeds in the domain. 

\begin{rem}
    The scheme \eqref{eqn:PC_Scheme} with the flux function \eqref{eqn:numerical_flux} can be considered a path-conservative generalisation of the central-upwind Kurganov-Tadmor scheme \cite{KurganovTadmor:2000}. In particular, by repeating the derivation from Sections 2-3 of Diaz et al.\ \cite{DiazKurganovdeLuna:2019} with the Kurganov-Tadmor flux \eqref{eqn:numerical_flux} it is straightforward to obtain the scheme \eqref{eqn:PC_Scheme}.
\end{rem}

\subsection{Velocity Constraint Enforcement}
While evolving using the path-conservative scheme \eqref{eqn:PC_Scheme} significantly improves the stability and accuracy of our simulations,  numerical errors may still lead to the velocity constraint \eqref{eqn:spatial_fluid_constraint} being violated. To resolve this, we enforce the constraint \eqref{eqn:spatial_fluid_constraint} by rescaling the components of the fluid velocity at each point in space where $|\nu|\geq 1$, 
\begin{align}
\label{eqn:velocity_constraint_rescale}
    \nu^{A}_{\text{new}} = \frac{\nu^{A}_{\text{old}}(1-\epsilon)}{|\nu|_{\text{old}}}.
\end{align}
The parameter $\epsilon$ is a small number typically taken to be $10^{-13}$ and the re-scaling \eqref{eqn:velocity_constraint_rescale} is implemented at the end of each individual Runge-Kutta substep (i.e. four times per time step for our fourth order integrator). While this is, admittedly, an ad-hoc way of enforcing \eqref{eqn:spatial_fluid_constraint}, we have found it to be remarkably robust. \newline \par

We have also investigated enforcing \eqref{eqn:spatial_fluid_constraint} using a physical-constraint-preserving scheme, see for example \cites{ZhangShu:2010,ZhangShu:2011,Wu:2017}, which enforces the constraint via a time step restriction. In practice, we have found this restriction leads to time step sizes of $\Delta t \leq 10^{-8}$ and often as small as $10^{-15}$. This appears to be because the fluid can be close to extremely tilted (i.e. $|\nu|\approx 1$) somewhere in the domain for essentially the entire simulation. Thus, to ensure the constraint \eqref{eqn:spatial_fluid_constraint} is satisfied, the most severe time step restriction must be enforced for almost the entire simulation. As such, these schemes do not seem to be a viable option for long simulations such as the ones presented in this article. 

\subsection{Initial Data}
 In this article, we focus on the following family of initial data at $t=0$,
\begin{equation}
\begin{gathered}
\label{eqn:numerical_initial_data}
    \nu^{1} = c\sin(x), \quad 
    \nu^{2} = \frac{-3ac}{2\sqrt{3}}\sin(x), \\  
    \nu^{3} = \frac{-3bc}{2\sqrt{3}}\sin(x), \quad 
    \mu = 1, \quad \Sigma_{3} = a, \\ 
    \Sigma_{2} = b, \quad
    \Sigma_{-} = d\sin(x), \quad\Sigma_{\times} = d\cos(x), \\
    N_{\times} = N_{-} = 0, \quad E_{1}^{1} = 1,
\end{gathered}
\end{equation}
where $a$, $b$, $c$, $d$ are freely specifiable real numbers. For any choice of these parameters, the constraints \eqref{eqn:CM2_ArealGauge}-\eqref{eqn:CM3_ArealGauge} are initially satisfied to machine precision. In the case where $a=b=c=d=0$, we recover FLRW initial data. Thus, this choice of initial data can be interpreted as a non-linear perturbation of an FLRW spacetime. All plots in this article were generated with the parameter values $a=b=-1$, $c=0.01$, $d=0.1$.

\subsection{Convergence Tests}
We have verified the accuracy of our code with convergence tests using resolutions of $N = 500$, $1000$, $2000$, and $4000$ cells. The numerical discretisation error $\Delta$ is estimated by taking the $\log_{2}$ of the absolute value of the difference between each simulation and the highest resolution run. The differences are computed by taking the average of the cells in the highest resolution run which are contained in cells of the lower resolution runs\footnote{Note, this is the same conservative restriction process used in adaptive mesh refinement schemes for finite volume methods, see for example \cite{BergerColella:1989}*{\S 3}.} and subtracting the result from the coarse cell values. \newline \par

\begin{figure}[htbp]
\centering
\subfigure[Subfigure 1 list of figures text][$\Sigma_{-}$]{
\includegraphics[width=0.45\textwidth]{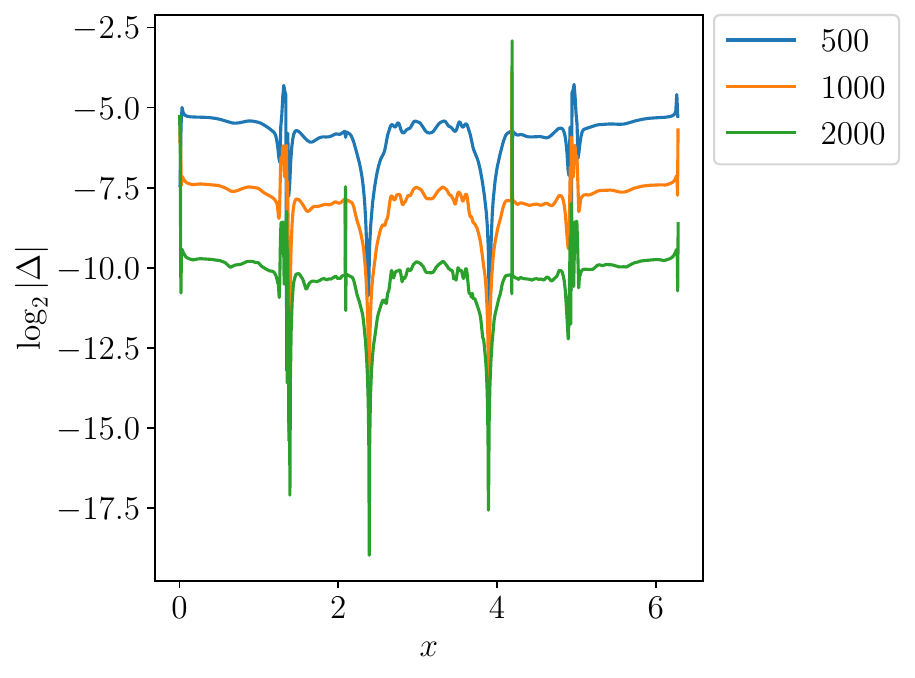}
\label{fig:Sigma_Minus_Convergence}}
\subfigure[Subfigure 2 list of figures text][$\nu^{1}$]{
\includegraphics[width=0.45\textwidth]{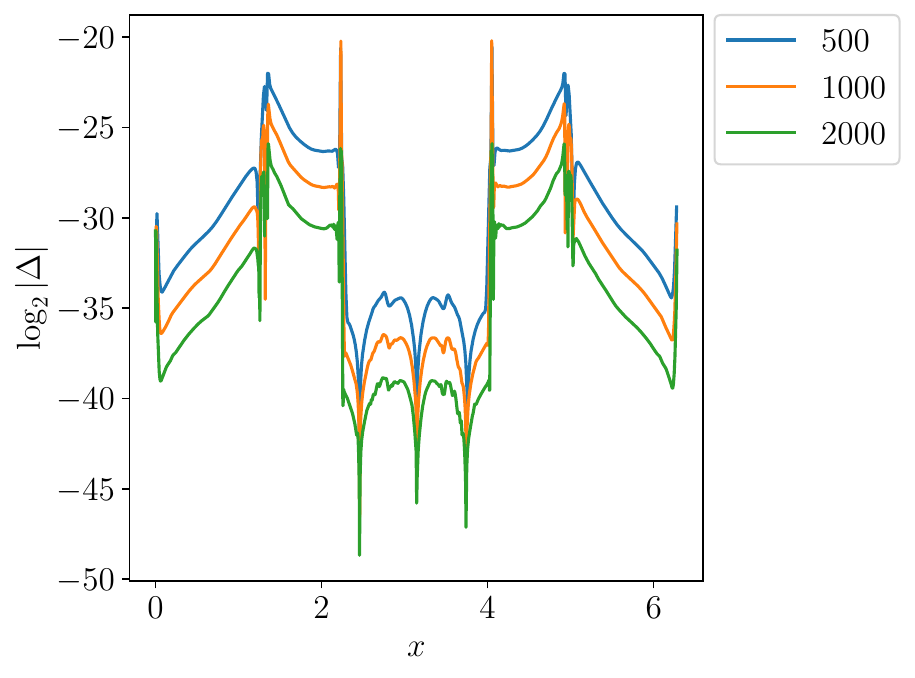}
\label{fig:nu1_convergence}}
\caption{Convergence plots of $\Sigma_{-}$ and $\nu^{1}$ at $t \approx -20$, $K=0.5$.}
\end{figure}

During the initial inhomogeneous part of the evolution, before the mixmaster regime, we observe approximately second-order convergence for all variables and constraints away from discontinuities. Near discontinuities, the convergence degrades to approximately first order. This is shown for $\Sigma_{-}$ and $\nu^{1}$ in Figures \ref{fig:Sigma_Minus_Convergence} and \ref{fig:nu1_convergence}, respectively. \newline \par 

\begin{figure}[htbp]
    \centering
    \includegraphics[width=0.5\linewidth]{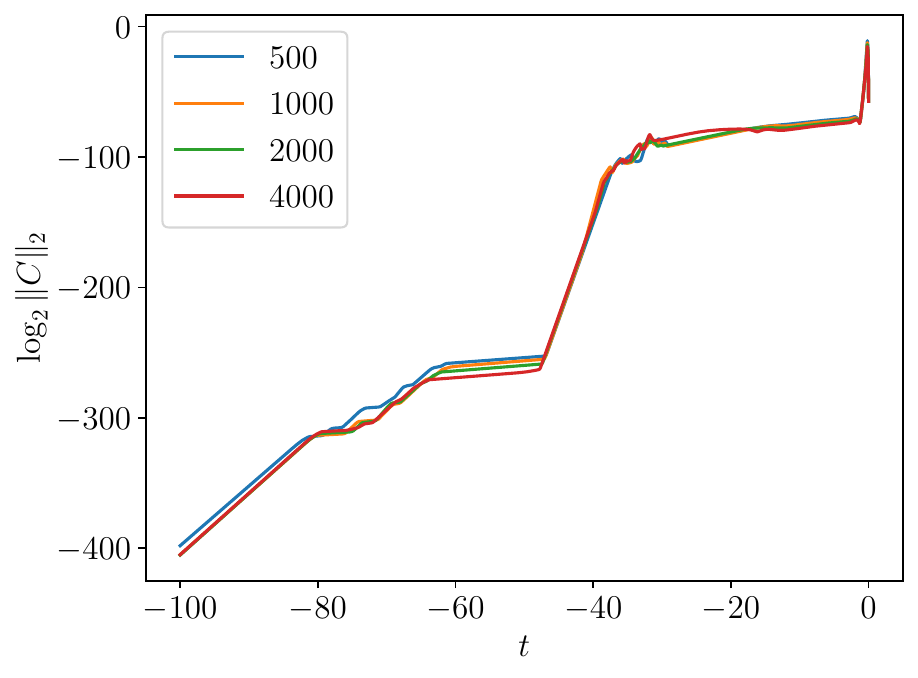}
    \caption{Convergence plots of $\log_{2}\big(\|(\mathcal{C}_{M})_{1}\|_{2}\big)$, $K=0.5$.}
    \label{fig:CM1_Convergence}
\end{figure}

As the chaotic mixmaster regime is reached, however, the convergence is notably worse. This is to be expected; numerical simulations of chaotic systems using standard techniques will not, in general, recover convergence to a reference solution. Nonetheless, as in \cite{Garfinkle:2004}, we observe that the constraint violation remains small during this regime, shown for the constraint $(\mathcal{C}_{M})_{1}$ in Figure \ref{fig:CM1_Convergence}. Overall, the numerical scheme appears to be robust and, as discussed below, produces results which closely agree with the spatially homogeneous heuristics. 

\section{Numerical Results}
\label{sec:Numerical_Results}
\begin{figure}[htbp]
\centering
\subfigure[Subfigure 1 list of figures text][$t=0.0$]{
\includegraphics[width=0.4\textwidth]{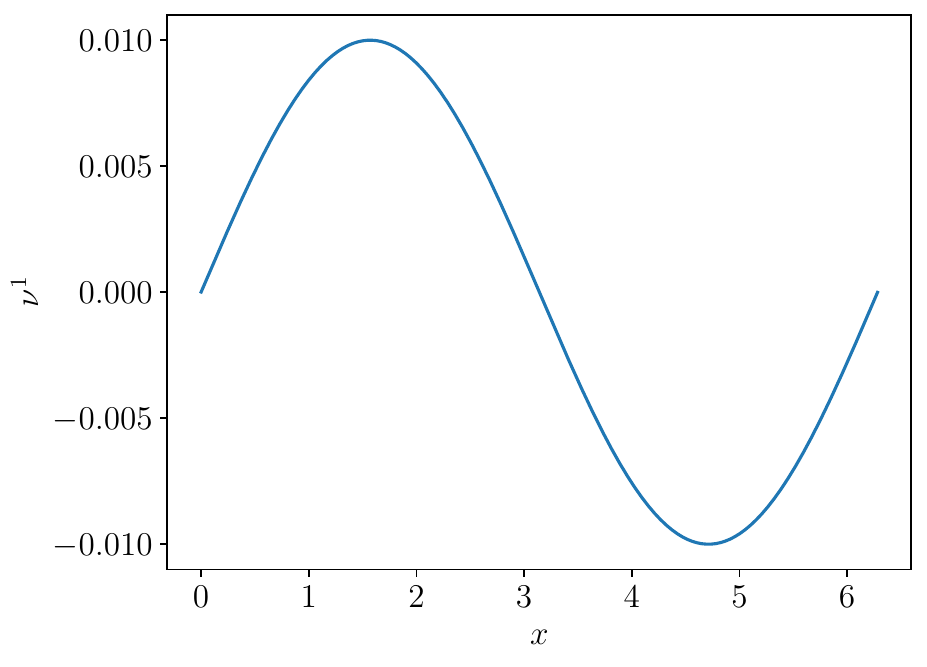}
\label{fig:nu1_shock_t0}}
\subfigure[Subfigure 2 list of figures text][$t=-0.25$]{
\includegraphics[width=0.4\textwidth]{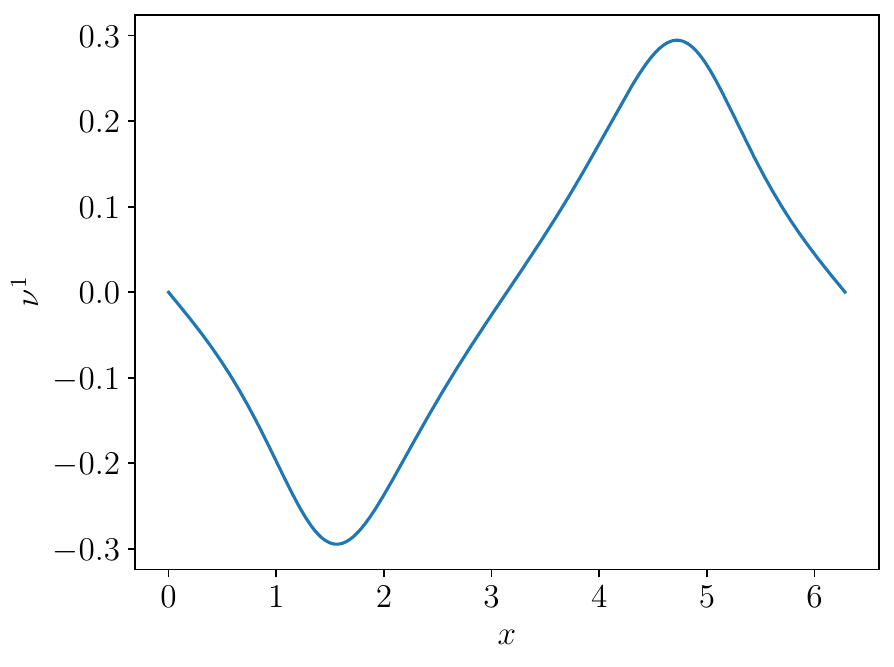}
\label{fig:nu1_shock_t025}}
\subfigure[Subfigure 2 list of figures text][$t=-0.5$]{
\includegraphics[width=0.4\textwidth]{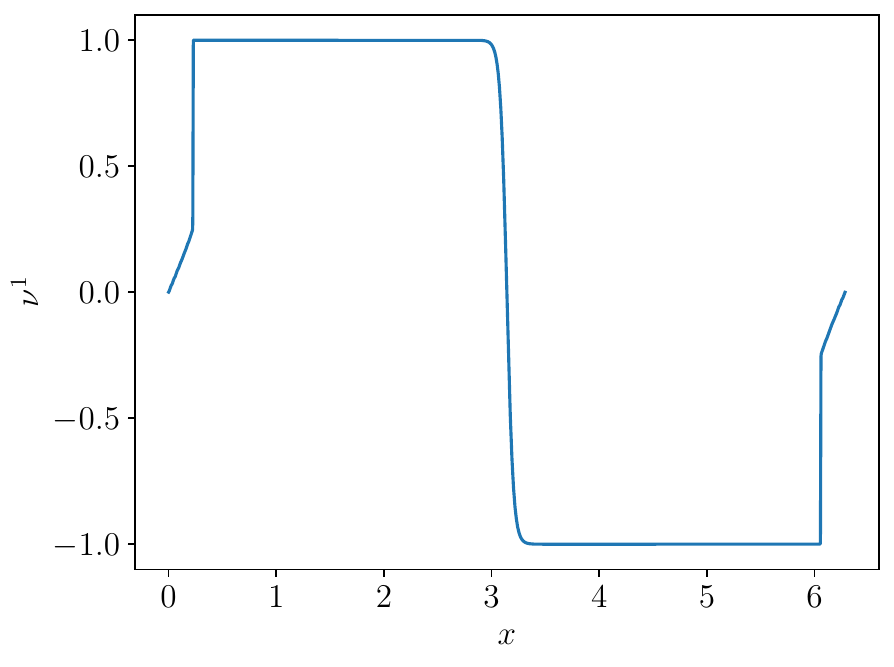}
\label{fig:nu1_shock_t05}}
\subfigure[Subfigure 2 list of figures text][$t=-1.0$]{
\includegraphics[width=0.4\textwidth]{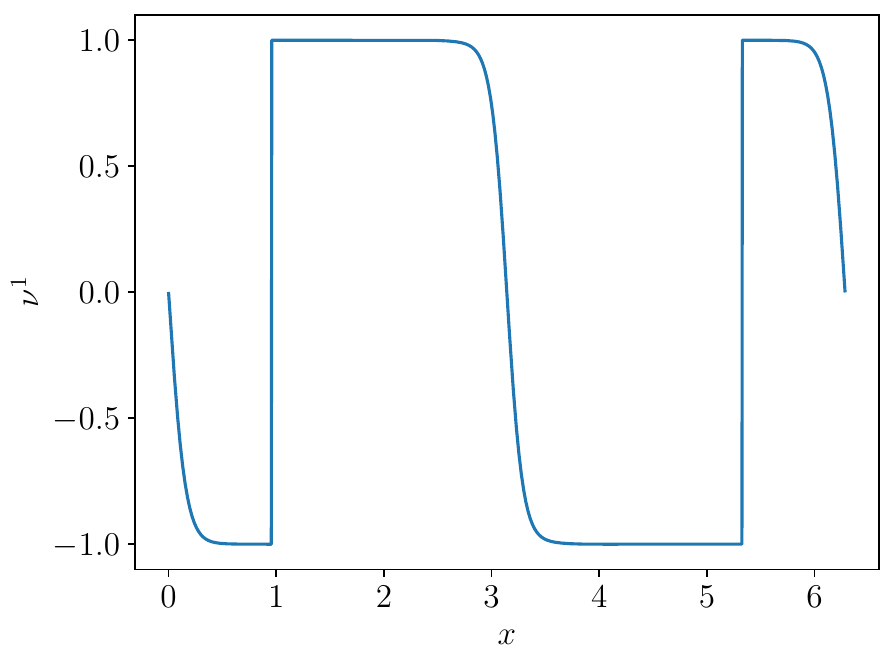}
\label{fig:nu1_shock_t1}}
\caption{Shock formation in $\nu^{1}$. $N=4000$, $K=0.5$.}
\label{fig:nu_shocks}
\end{figure}

Let us now discuss the behaviour of a typical simulation. Initially the evolution proceeds in an inhomogeneous manner and shocks rapidly develop in the fluid velocity $\nu^{A}$, shown for the component $\nu^{1}$ in Figure \ref{fig:nu_shocks}. In particular, we consistently find that $\nu^{1}$ almost immediately approaches an extreme tilt (i.e. $\nu^{1} \approx \pm 1$) while the other components of the fluid velocity remain small. \newline \par

\begin{figure}[htbp]
    \centering
    \includegraphics[width=0.5\linewidth]{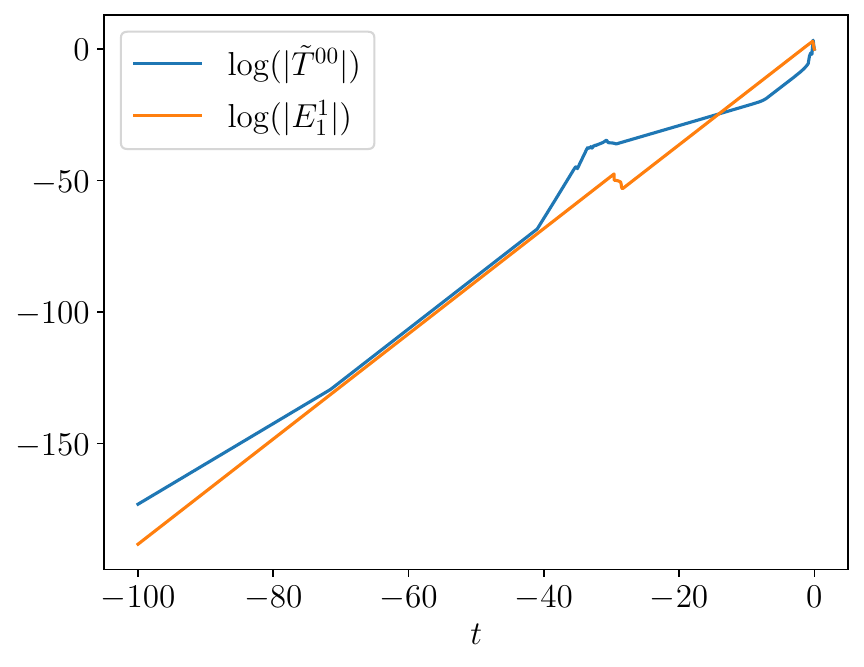}
    \caption{Maximum values of $\log(|\tilde{T}^{00}|)$ and $\log(|E_{1}^{1}|)$ over time. $N=4000$, $K=0.5$.}
    \label{fig:E11_T00_decay}
\end{figure}

The BKL conjecture predicts that the spatial derivatives and matter should become negligible as the singularity is approached (cf. Section \ref{sec:Homogeneous_Asymptotics}). Specifically, we expect that both the spatial frame $E_{1}^{1}$ and stress-energy components $\tilde{T}^{ab}$ should decay to zero. We have tested this by monitoring the size of $\log(|\tilde{T}^{00}|)$ and $\log(|E^{1}_{1}|)$ over the course of the evolution. As expected, these variables exponentially decay as $t\rightarrow -\infty$, shown in Figure \ref{fig:E11_T00_decay}, and by $t \approx -28$ the oscillatory mixmaster dynamics begin. \newline \par

\subsection{Tilt Transitions}
\begin{figure}[htbp]
    \centering
    \includegraphics[width=0.5\linewidth]{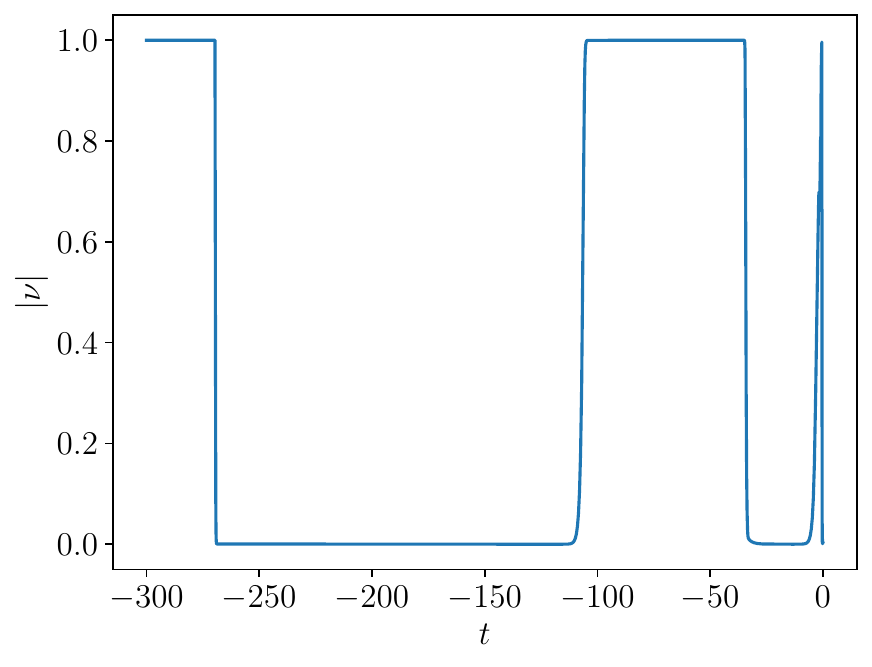}
    \caption{Tilt transitions in $|\nu|$ at single point in space. $N=500$, $K=0.5$.}
    \label{fig:tilt_transitions_nu_norm}
\end{figure}

Since the stress-energy components are negligible, the evolution of the gravitational variables is indistinguishable from previous studies in vacuum, see for example \cites{BergerIsenbergWeaver:2001,Garfinkle:2004,Andersson_et_al:2005}. Thus, for the remainder of this article we will focus on the behaviour of the fluid. In Figure \ref{fig:tilt_transitions_nu_norm}, the norm of the spatial fluid velocity $|\nu|$ at a single (typical) point in space is plotted. In particular, we observe that the fluid oscillates between orthogonal and extremely tilted states, demonstrating that tilt transitions \textit{can} occur in inhomogeneous cosmologies. \newline \par

\begin{figure}[htbp]
    \centering
    \includegraphics[width=0.8\linewidth]{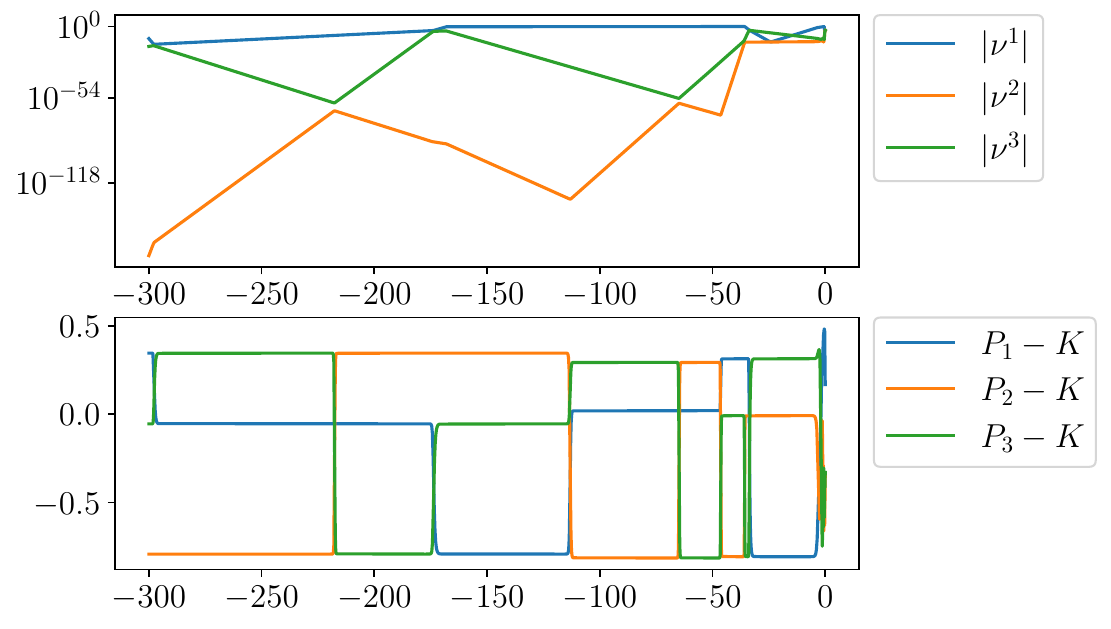}
    \caption{Comparison of $\nu^{A}$ and $P_{A}-K$ at the $520^{\text{th}}$ cell. Observe that changes in the growth and decay of the velocity components coincide with the changes in sign and size of the quantities $P_{A}-K$, following the trigger rules in Table \ref{tab:velocity_triggers}. $N=1000$, $K=0.5$.}
    \label{fig:tilt_transitions_nu_components_trigger_comparison}
\end{figure}

\begin{figure}[htbp]
    \centering
    \includegraphics[width=0.5\linewidth]{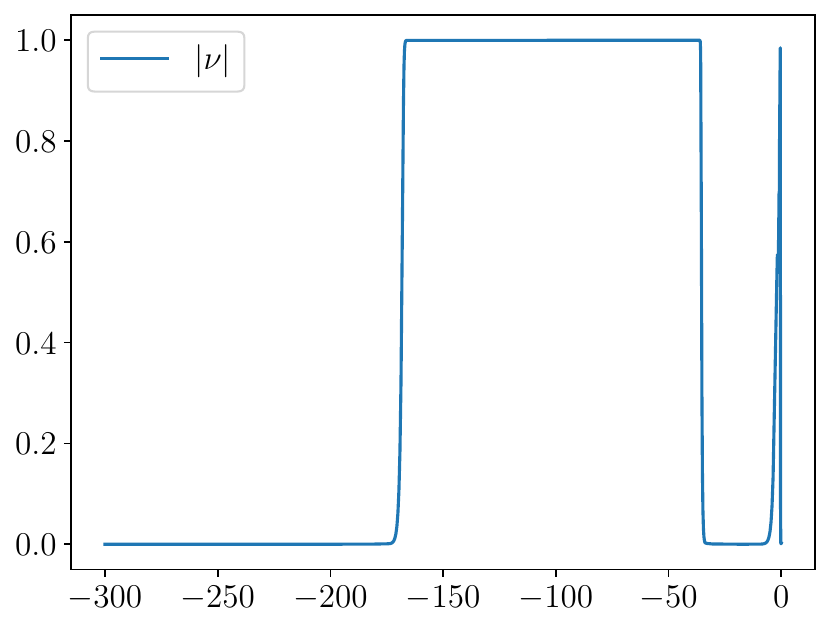}
    \caption{The value of the norm $|\nu|$ at the $520^{\text{th}}$ cell. $N=1000$, $K=0.5$.}
    \label{fig:tilt_transitions_nu_norm_trigger_comparison}
\end{figure}

Next, we want to determine whether the tilt transitions are, in fact, caused by the trigger conditions derived from our linearised analysis in Section \ref{sec:Homogeneous_Asymptotics}. In Figure \ref{fig:tilt_transitions_nu_components_trigger_comparison}, we compare the change in the components $|\nu^{A}|$ with the quantities $P_{A}-K$ at a single point. The value of the norm $|\nu|$ at the same point is shown in Figure \ref{fig:tilt_transitions_nu_norm_trigger_comparison}. From these plots, it is clear that the trigger conditions on the Kasner exponents in Table \ref{tab:velocity_triggers} do correspond to changes in the growth/decay of the velocity components. \newline \par 

In practice, we find that the components $\nu^{2}$ and $\nu^{3}$ rapidly decay to zero and the tilt transitions almost exclusively occur in the $\nu^{1}$ component of the fluid. Even when a trigger for a tilt transition in $\nu^{2}$ or $\nu^{3}$ is active, these components grow too slowly to noticeably affect the evolution and remain less than the threshold of numerical precision (approx $10^{-15}$). This is clear by looking at the component $|\nu^{2}|$ in Figure \ref{fig:tilt_transitions_nu_components_trigger_comparison}. Even though the trigger for the component $\nu^{2}$ is active between approximately $-220 \leq t \leq -110$, the value of $|\nu^{2}|$ never rises above $10^{-54}$. It is possible this is a consequence of the $\mathbb{T}^{2}$-symmetry and that general $3+1$ evolutions will consistently have tilt transitions in all components of the fluid velocity. We intend to investigate this in a future work. \newline \par

\begin{figure}[htbp]
    \centering
    \includegraphics[width=0.8\linewidth]{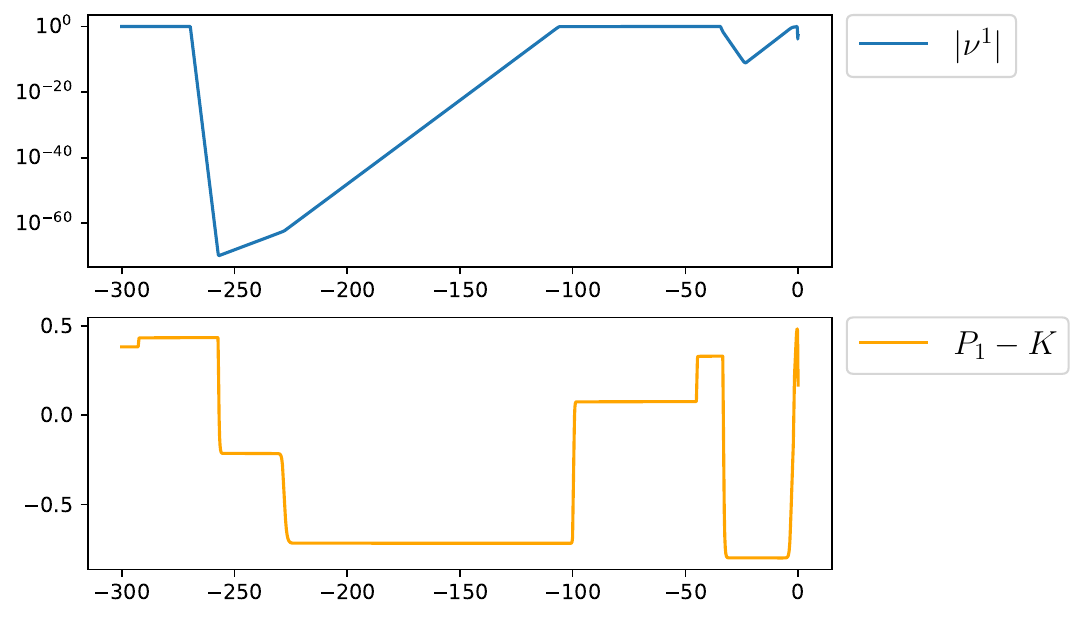}
    \caption{Comparison of $|\nu^{1}|$ and $P_{1}-K$ at a single point in space. $N=500$, $K=0.5$.}
    \label{fig:tilt_transitions_nu1_only}
\end{figure}

Now, since $\nu^{2}$ and $\nu^{3}$ are almost always close to zero, the trigger conditions in Table \ref{tab:velocity_triggers} essentially reduce to a condition on the sign of $P_{1}-K$. That is, $\nu^{1}$ grows when $P_{1}-K>0$ and decays when $P_{1}-K<0$. Moreover, the growth/decay rate should increase with the size of $|P_{1}-K|$. This is clearly replicated by our numerical solutions, shown in Figure \ref{fig:tilt_transitions_nu1_only}. \newline \par 

\begin{figure}[htbp]
    \centering
    \includegraphics[width=0.6\linewidth]{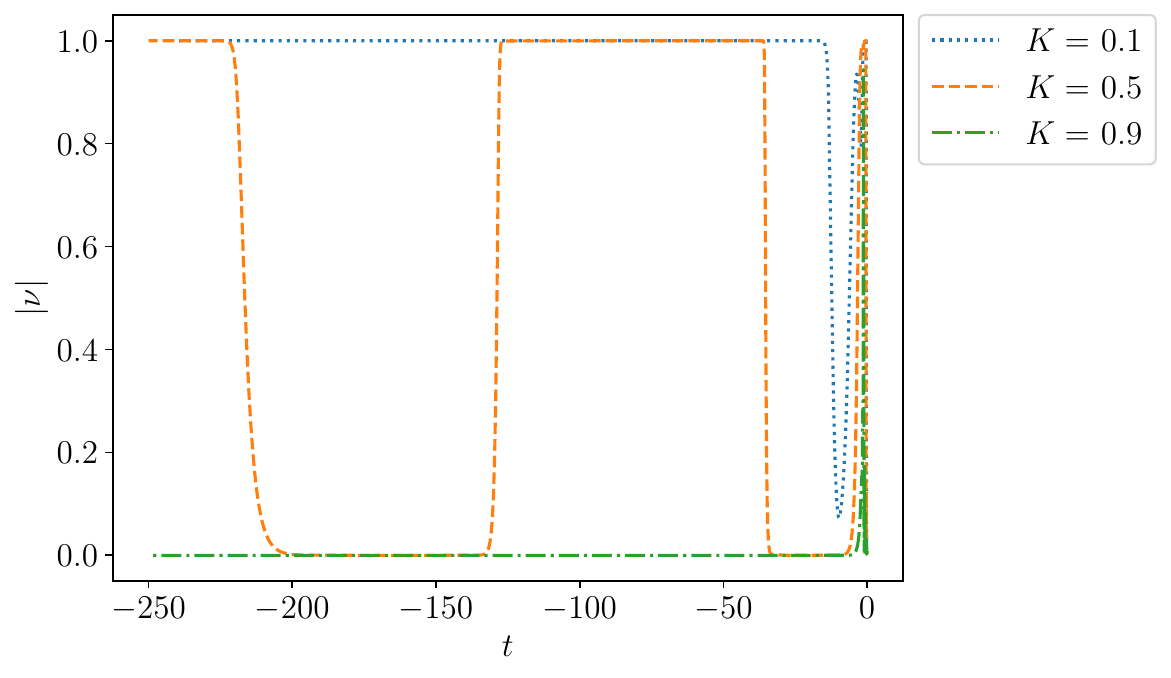}
    \caption{Comparison of tilt transitions in the same grid cell for different values of the sound speed parameter $K$. $N=500$}
    \label{fig:K_Comparison_Tilt}
\end{figure}

We also expect the size of the sound speed parameter $K$ to affect the frequency of tilt transitions. To see this, recall that the Kasner exponents must satisfy the conditions \eqref{eqn:Kasner_Conditions},  
\begin{align*}
    \sum_{A = 1}^{3}P_{A} = 1, \quad \sum_{A = 1}^{3}P_{A}^{2} = 1.
\end{align*}
In particular, the second condition implies $P_{1} \leq 1$ and, hence, 
\begin{align*}
    P_{1} - K \leq 1-K
\end{align*}
Thus, as $K \nearrow 1$, we expect that $P_{1}-K$ will tend to be negative which, by our trigger condition, will cause $\nu^{1}$ to decay to zero. Moreover, in the limiting case $K=1$, we have
\begin{align*}
    P_{1} - K \leq 0,
\end{align*}
which suggests that a stiff fluid must tend to an orthogonal state. This is consistent with both the numerical observations of Curtis and Garfinkle \cite{CurtisGarfinkle:2005} and the analytic results of Rodnianski and Speck \cites{RodnianskiSpeck:2018b,RodnianskiSpeck:2018c} on stiff fluid singularities. Conversely, as $K \searrow 0$, $P_{1}-K$ will tend to be positive which would cause the fluid to approach towards an extremely tilted state. This also suggests that somewhere in between these two extremes, the fluid will consistently oscillate between the orthogonal and extremely tilted states. Indeed, this heuristic argument is supported by our numerical results. In Figure \ref{fig:K_Comparison_Tilt}, we compare the evolution of $|\nu|$ for $K=0.1$, $0.5$, and $0.9$. As predicted, the $K=0.1$ and $K=0.9$ solutions tend to extremely tilted and orthogonal states, respectively, while the $K=0.5$ solution oscillates between the two states.

\subsection{Density Inhomogeneities}
Beginning with the influential work of Rendall \cite{Rendall:2004}, a series of previous studies \cites{ColeyLim:2012,ColeyLim:2013,ColeyLim:2015,BMO:2023,Oliynyk:2024,BMO:2024} have found that the transition between orthogonal and tilted behaviour at adjacent points in space causes inhomogeneities in the fluid density. In particular, the fractional density gradient $\frac{\del_{x}\mu}{\mu}$, which measures the inhomogeneity of the spacetime, develops large, spiky features and blows-up at these transition points - a phenomenon now known as the `Rendall instability'. \newline \par

\begin{figure}
    \centering
    \includegraphics[width=0.5\linewidth]{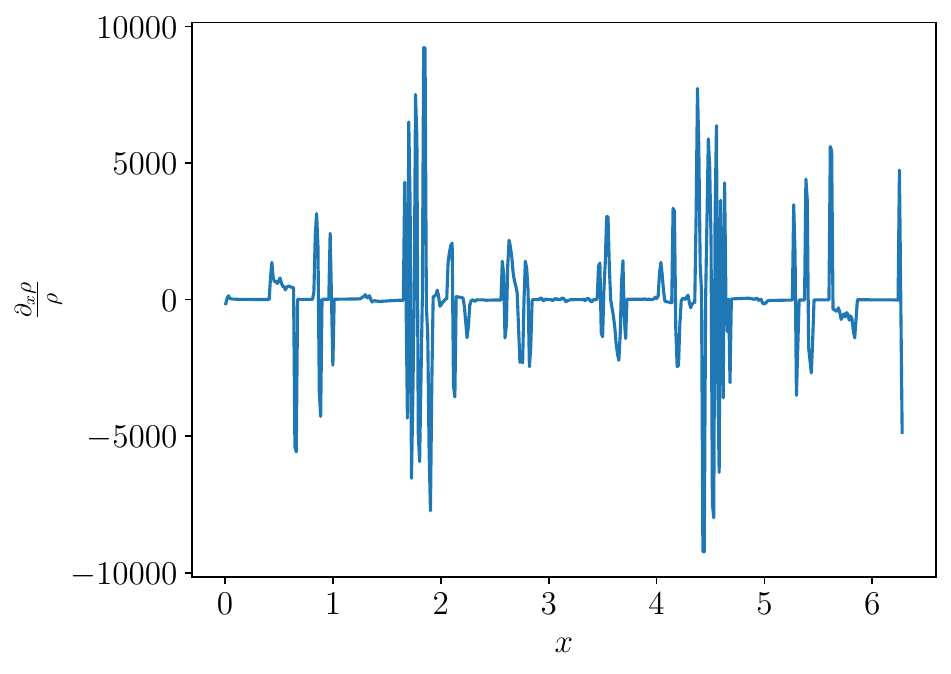}
    \caption{The density gradient of the fluid at $t\approx -290$, $N=500$.}
    \label{fig:density_gradient}
\end{figure}

Previous numerical studies of the Rendall instability have focused on Gowdy-symmetric solutions with $\mathbb{T}^{3}$ topology. This is a subclass of $\mathbb{T}^{2}$-symmetric spacetimes which does not display mixmaster behaviour near the big bang. For Gowdy-symmetric models, the fluid asymptotically approaches either an orthogonal or extremely tilted state but cannot not oscillate between the two. Typically, this means that density inhomogeneities are only generated around a few points in space. However, the oscillatory nature of the fluid near the big bang in $\mathbb{T}^{2}$-symmetry means that these density inhomogeneities can be generated all over the spatial domain in a chaotic fashion, shown in Figure \ref{fig:density_gradient}. This indicates, as previously suggested by Coley and Lim \cite{ColeyLim:2012}, that mixmaster oscillations are a possible mechanism for creating primordial density fluctuations in the early universe and could therefore play an important role in structure formation.  

\section{Discussion}
We have performed simulations of the approach to the big bang in $\mathbb{T}^{2}$-symmetric cosmological models containing a non-stiff perfect fluid. Our numerical implementation utilises a path-conservative finite volume scheme based on \cite{DiazKurganovdeLuna:2019} which allows us to evolve the primitive fluid variables directly while maintaining shock-capturing properties. This was essential for obtaining long-term stable evolutions. Our results demonstrate that the fluid velocity develops oscillatory, mixmaster-esque behaviour, known as tilt transitions, which are induced by oscillations in the gravitational variables. In particular, this provides convincing numerical support for the generalised BKL conjecture of Uggla et al.\ \cite{UEWE:2003}. Moreover, we find that these tilt transitions generate local inhomogeneities in the matter density, which may prove important for structure formation in the early universe. There are numerous interesting directions for future analytic and numerical work. From the numerical perspective, the obvious extension is to evolve models without symmetry assumptions. Indeed, we expect that the numerical scheme used here could be extended in a straightforward manner to 3+1 simulations. Additionally, our path-conservative approach for evolving the fluid may prove useful in other situations where the standard conservative variables have problematic numerical behaviour. A natural first step for analytic work would be to rigorously prove a past attractor theorem for tilted Bianchi II cosmologies, analogous to Ringstrom's proof for Bianchi IX models \cite{Ringstrom:2001}. Similarly, in the spatially inhomogeneous setting, one could try to generalise the recent results of Li \cites{Li:2024a,Li:2024b} to fluid models. Finally, at least some portion of singularities in asymptotically flat spacetimes is expected to be described by the BKL picture. It would be interesting to extend the methods presented here to this setting as well.

\bibliography{refs.bib}

@article{belinskii1970,
	author = {Belinskii, V.A. and Khalatnikov, I.M. and Lifshitz, E.M.},
	doi = {10.1080/00018737000101171},
	journal = {Adv. Phys.},
	number = {80},
	pages = {525--573},
	title = {{O}scillatory {A}pproach to a {S}ingular {P}oint in {R}elativistic {C}osmology},
	volume = {19},
	year = {1970}}

@article{BergerIsenbergWeaver:2001,
	author = {Berger, B.K. and Isenberg, J. and Weaver, M.},
	journal = {Phys. Rev. D},
	number = {8},
	title = {{{O}scillatory {A}pproach to the {S}ingularity in {V}acuum {S}pacetimes with $T^2$ {I}sometry}},
	volume = {64},
	year = {2001}}

@article{BMO:2023,
	author = {F. Beyer and E. Marshall and T.A. Oliynyk},
	doi = {10.1103/PhysRevD.107.104030},
	journal = {Phys. Rev. D},
	pages = {104030},
	publisher = {American Physical Society},
	title = {{F}uture {I}nstability of {FLRW} {F}luid {S}olutions for {L}inear {E}quations of {S}tate {$p=K\rho$} with {$1/3<K<1$}},
	url = {https://link.aps.org/doi/10.1103/PhysRevD.107.104030},
	volume = {107},
	year = {2023}}

@article{BeyerOliynyk:2023,
   title={{L}ocalized {B}ig {B}ang {S}tability for the {E}instein-{S}calar {F}ield {E}quations},
   volume={248},
   ISSN={1432-0673},
   url={http://dx.doi.org/10.1007/s00205-023-01939-9},
   DOI={10.1007/s00205-023-01939-9},
   number={1},
   journal={Archive for Rational Mechanics and Analysis},
   publisher={Springer Science and Business Media LLC},
   author={Beyer, F. and Oliynyk, T.A.},
   year={2023},
   month=dec }

@article{BOZ:2025,
      title={{L}ocalized {P}ast {S}tability of the {S}ubcritical {K}asner-{S}calar {F}ield {S}pacetimes}, 
      author={F. Beyer and T.A. Oliynyk and W. Zheng},
      year={2025},
      eprint={2502.09210},
      archivePrefix={arXiv},
      primaryClass={gr-qc},
      url={https://arxiv.org/abs/2502.09210}, 
}

@article{ColeyLim:2015,
	author = {Coley, A.A. and Lim, W.C.},
	doi = {10.1088/0264-9381/33/1/015009},
	issn = {0264-9381, 1361-6382},
	journal = {Classical and Quantum Gravity},
	number = {1},
	pages = {015009},
	title = {{S}pikes and {M}atter {I}nhomogeneities in {M}assless {S}calar {F}ield {M}odels},
	volume = {33},
	year = {2016}}

@article{CurtisGarfinkle:2005,
	author = {Curtis, J. and Garfinkle, D.},
	doi = {10.1103/PhysRevD.72.064003},
	issn = {1550-7998, 1550-2368},
	journal = {Physical Review D},
	number = {6},
	pages = {064003},
	title = {Numerical {S}imulations of {S}tiff {F}luid {G}ravitational {S}ingularities},
	volume = {72},
	year = {2005}}

@article{Fournodavlos_et_al:2023,
	author = {Fournodavlos, G. and Rodnianski, I. and Speck, J.},
	doi = {10.1090/jams/1015},
	issn = {0894-0347, 1088-6834},
	journal = {J. Amer. Math. Soc.},
	month = jul,
	pages = {827--916},
	shorttitle = {Stable {Big} {Bang} formation for {Einstein}'s equations},
	title = {{S}table {Big} {Bang} {F}ormation for {Einstein}'s {E}quations: {The} {C}omplete {S}ub-{C}ritical {R}egime},
	url = {https://www.ams.org/jams/2023-36-03/S0894-0347-2023-01015-X/},
	urldate = {2023-08-03},
	volume = {36},
	year = {2023}}

@article{Garfinkle:2004,
	author = {Garfinkle, D.},
	doi = {10.1103/PhysRevLett.93.161101},
	journal = {Physical Review Letters},
	number = {16},
	pages = {6},
	title = {Numerical {{Simulations}} of {{Generic Singularities}}},
	volume = {93},
	year = {2004}}

@article{Garfinkle:2007,
	author = {Garfinkle, D.},
	doi = {10.1088/0264-9381/24/12/S19},
	issn = {0264-9381, 1361-6382},
	journal = {Classical and Quantum Gravity},
	number = {12},
	pages = {S295-S306},
	title = {Numerical Simulations of General Gravitational Singularities},
	volume = {24},
	year = {2007}}

@book{HawkingEllis:1973,
	author = {Hawking, S.W. and Ellis, G.F.R.},
	doi = {10.1017/CBO9780511524646},
	edition = {First},
	isbn = {978-0-521-09906-6 978-0-521-20016-5 978-0-511-52464-6},
	publisher = {{Cambridge University Press}},
	title = {{T}he {L}arge {S}cale {S}tructure of {S}pace-{T}ime},
	year = {1973}}

@article{Lim:2009,
	author = {Lim, W.C. and Andersson, L. and Garfinkle, D. and Pretorius, F.},
	doi = {10.1103/PhysRevD.79.123526},
	journal = {Physical Review D},
	number = {12},
	pages = {123526},
	title = {{S}pikes in the {M}ixmaster {R}egime of {$G_2$} {C}osmologies},
	volume = {79},
	year = {2009}}

@article{Rendall:2004,
	author = {Rendall, A.D.},
	doi = {10.1007/s00023-004-0189-1},
	journal = {Ann. Henri Poincar{\'e}},
	number = {6},
	pages = {1041-1064},
	title = {{A}symptotics of {S}olutions of the {{Einstein}} {E}quations with {P}ositive {C}osmological {C}onstant},
	volume = {5},
	year = {2004}}

@article{RodnianskiSpeck:2018b,
	author = {I.~Rodnianski and J.~Speck},
	doi = {10.4007/annals.2018.187.1.2},
	issn = {0003-486X},
	journal = {Ann. Math.},
	number = {1},
	pages = {65--156},
	title = {{A} {R}egime of {L}inear {S}tability for the {{Einstein}}-{S}calar {F}ield {S}ystem with {A}pplications to {N}on-{L}inear {{Big Bang}} {F}ormation},
	volume = {187},
	year = {2018}}

@article{RodnianskiSpeck:2018c,
	author = {I.~Rodnianski and J.~Speck},
	doi = {10.1007/s00029-018-0437-8},
	issn = {1022-1824, 1420-9020},
	journal = {Sel. Math. New Ser.},
	number = {5},
	pages = {4293--4459},
	title = {{S}table {{Big Bang}} {F}ormation in {N}ear-{{FLRW}} {S}olutions to the {{Einstein}}-{S}calar {F}ield and {{Einstein}}-{S}tiff {F}luid {S}ystems},
	volume = {24},
	year = {2018}}

@article{Ringstrom:2001,
	author = {H. Ringstr{\"o}m},
	journal = {Ann. Henri Poincar\'{e}},
	pages = {405-500},
	title = {The {B}ianchi {IX} Attractor},
	volume = {2},
	year = {2001}}

@article{andersson2001,
	author = {Andersson, L. and Rendall, A.D.},
	doi = {10.1007/s002200100406},
	journal = {Commun. Math. Phys.},
	number = {3},
	pages = {479-511},
	title = {Quiescent {{Cosmological Singularities}}},
	volume = {218},
	year = {2001}}

@article{BergerGarfinkle:1998,
	author = {Berger, B.K. and Garfinkle, D.},
	doi = {10.1103/PhysRevD.57.4767},
	journal = {Physical Review D},
	note = {DOI:~\href{https://doi.org/10.1103/PhysRevD.57.4767}{10.1103/PhysRevD.57.4767}},
	number = {8},
	pages = {4767--4777},
	title = {Phenomenology of the {{Gowdy}} Universe on {$T^3\times R$}},
	volume = {57},
	year = {1998}}

@article{LeFlochRendall:2011,
	author = {LeFloch, P.G. and Rendall, A.D.},
	date = {2011},
	doi = {10.1007/s00205-011-0425-z},
	id = {LeFloch2011},
	isbn = {1432-0673},
	journal = {Arch. Rat. Mech.},
	number = {3},
	pages = {841--870},
	title = {{A} {G}lobal {F}oliation of {E}instein-{E}uler {S}pacetimes with {G}owdy-{S}ymmetry on {$T^3$}},
	url = {https://doi.org/10.1007/s00205-011-0425-z},
	volume = {201},
	year = {2011}}

@article{UEWE:2003,
   title={Past {A}ttractor in {I}nhomogeneous {C}osmology},
   volume={68},
   ISSN={1089-4918},
   url={http://dx.doi.org/10.1103/PhysRevD.68.103502},
   DOI={10.1103/physrevd.68.103502},
   number={10},
   journal={Physical Review D},
   publisher={American Physical Society (APS)},
   author={Uggla, C. and van Elst, H. and Wainwright, J. and Ellis, G.F.R.},
   year={2003},
   month=nov }

@article{Elst_et_al:2001,
   title={{D}ynamical {S}ystems {A}pproach to {$G_{2}$} {C}osmology},
   volume={19},
   ISSN={1361-6382},
   url={http://dx.doi.org/10.1088/0264-9381/19/1/304},
   DOI={10.1088/0264-9381/19/1/304},
   number={1},
   journal={Classical and Quantum Gravity},
   publisher={IOP Publishing},
   author={van Elst, H. and Uggla, C. and Wainwright, J.},
   year={2001},
   month=dec, pages={51–82} }

@book{EllisWainwright:1997,
  title={{D}ynamical {S}ystems in {C}osmology},
  author={Wainwright, J. and Ellis, G.F.R.},
  isbn={9780521554572},
  lccn={96028592},
  year={1997},
  publisher={Cambridge University Press}
}

@article{Andersson_et_al:2005,
  title = {{A}symptotic {S}ilence of {G}eneric {C}osmological {S}ingularities},
  author = {Andersson, L. and van Elst, H. and Lim, W.C. and Uggla, C.},
  journal = {Phys. Rev. Lett.},
  volume = {94},
  issue = {5},
  pages = {051101},
  numpages = {4},
  year = {2005},
  publisher = {American Physical Society},
  doi = {10.1103/PhysRevLett.94.051101},
  url = {https://link.aps.org/doi/10.1103/PhysRevLett.94.051101}
}

@book{LeVeque:2002, 
place={Cambridge}, 
series={Cambridge Texts in Applied Mathematics}, 
title={Finite {V}olume {M}ethods for {H}yperbolic {P}roblems}, publisher={Cambridge University Press}, 
author={LeVeque, R.J.}, 
year={2002}, collection={Cambridge Texts in Applied Mathematics}}

@article{KurganovTadmor:2000,
  title={{N}ew {H}igh-{R}esolution {C}entral {S}chemes for {N}on-{L}inear {C}onservation {L}aws and {C}onvection-{D}iffusion {E}quations},
  author={Kurganov, A. and Tadmor, E.},
  journal={Journal of computational physics},
  volume={160},
  number={1},
  pages={241--282},
  year={2000},
  publisher={Elsevier}
}

@article{BMO:2024,
   title={Past {I}nstability of {FLRW} {S}olutions of the {E}instein-{E}uler-{S}calar {F}ield {E}quations for {L}inear {E}quations of {S}tate {$p=K\rho$} with {$0\leq K<1/3$}},
   volume={110},
   ISSN={2470-0029},
   url={http://dx.doi.org/10.1103/PhysRevD.110.044060},
   DOI={10.1103/physrevd.110.044060},
   number={4},
   journal={Physical Review D},
   publisher={American Physical Society (APS)},
   author={Beyer, F. and Marshall, E. and Oliynyk, T.A.},
   year={2024}}

@article{ElstUggla:1997,
   title={General {R}elativistic {O}rthonormal {F}rame {A}pproach},
   volume={14},
   ISSN={1361-6382},
   url={http://dx.doi.org/10.1088/0264-9381/14/9/021},
   DOI={10.1088/0264-9381/14/9/021},
   number={9},
   journal={Classical and Quantum Gravity},
   publisher={IOP Publishing},
   author={van Elst, H. and Uggla, C.},
   year={1997},
   month=sep, pages={2673–2695} }

@article{MacCallum:1973,
    author = {MacCallum, M.A.H.},
    editor = {Schatzman, E.},
    title = {Cosmological {M}odels from a {G}eometric {P}oint of {V}iew},
    eprint = {2001.11387},
    archivePrefix = {arXiv},
    journal = {Cargese Lect. Phys.},
    volume = {6},
    pages = {61-174},
    year = {1973},
    primaryClass={gr-qc},
      url={https://arxiv.org/abs/2001.11387}
}

@article{EllisMaccallum:1969,
author = {G.F.R. Ellis and M.A.H. MacCallum},
title = {{A {C}lass of {H}omogeneous {C}osmological {M}odels}},
volume = {12},
journal = {Communications in Mathematical Physics},
number = {2},
publisher = {Springer},
pages = {108 -- 141},
year = {1969},
}

@article{Marshall:2025,
  title={{I}nstability of {S}lowly {E}xpanding {FLRW} {S}pacetimes},
  author={Marshall, E.},
  journal={Classical and Quantum Gravity},
  volume={42},
  number={10},
  pages={105005},
  year={2025},
  publisher={IOP Publishing}
}

@article{Fournodavlos_et_al:2024,
      title={{F}uture {S}tability of {P}erfect {F}luids with {E}xtreme {T}ilt and {L}inear {E}quation of {S}tate $p=c_s^2\rho$ for the {E}instein-{E}uler {S}ystem with {P}ositive {C}osmological {C}onstant: {T}he {R}ange $\frac{1}{3}<c_s^2<\frac{3}{7}$}, 
      author={G. Fournodavlos and E. Marshall and T.A. Oliynyk},
      year={2024},
      eprint={2404.06789},
      archivePrefix={arXiv},
      primaryClass={math.AP},
      url={https://arxiv.org/abs/2404.06789}}

@article{Hewitt_et_al:2001,
   title={{T}he {A}symptotic {R}egimes of {T}ilted {B}ianchi {II} {C}osmologies},
   volume={33},
   ISSN={1572-9532},
   url={http://dx.doi.org/10.1023/A:1002075902953},
   DOI={10.1023/a:1002075902953},
   number={1},
   journal={General Relativity and Gravitation},
   publisher={Springer Science and Business Media LLC},
   author={Hewitt, C.G. and Bridson, R. and Wainwright, J.},
   year={2001},
   month=jan, pages={65–94} }

@article{GarfinklePretorius:2020,
   title={{S}pike {B}ehavior in the {A}pproach to {S}pacetime {S}ingularities},
   volume={102},
   ISSN={2470-0029},
   url={http://dx.doi.org/10.1103/PhysRevD.102.124067},
   DOI={10.1103/physrevd.102.124067},
   number={12},
   journal={Physical Review D},
   publisher={American Physical Society (APS)},
   author={Garfinkle, D. and Pretorius, F.},
   year={2020},
   month=dec }

@article{Weaver_et_al:1998,
   title={Mixmaster {B}ehavior in {I}nhomogeneous {C}osmological {S}pacetimes},
   volume={80},
   ISSN={1079-7114},
   url={http://dx.doi.org/10.1103/PhysRevLett.80.2984},
   DOI={10.1103/physrevlett.80.2984},
   number={14},
   journal={Physical Review Letters},
   publisher={American Physical Society (APS)},
   author={Weaver, M. and Isenberg, J. and Berger, B.K.},
   year={1998},
   month=apr, pages={2984–2987} }

@article{BeyerOliynyk:2024,
   title={{P}ast {S}tability of {FLRW} {S}olutions to the {E}instein-{E}uler-{S}calar {F}ield {E}quations and their {B}ig {B}ang {S}ingularities},
   volume={1},
   ISSN={2994-8452},
   url={http://dx.doi.org/10.4310/BPAM.2024.v1.n2.a4},
   DOI={10.4310/bpam.2024.v1.n2.a4},
   number={2},
   journal={Beijing Journal of Pure and Applied Mathematics},
   publisher={International Press of Boston},
   author={Beyer, F. and Oliynyk, T.A.},
   year={2024},
   pages={515–637} }

@article{Groeniger_et_al:2023,
      title={{F}ormation of {Q}uiescent {B}ig {B}ang {S}ingularities}, 
      author={H.O. Groeniger and O. Petersen and H. Ringström},
      year={2023},
      eprint={2309.11370},
      archivePrefix={arXiv},
      primaryClass={gr-qc},
      url={https://arxiv.org/abs/2309.11370}, 
}

@article{Li:2024a,
      title={{BKL} {B}ounces {O}utside {H}omogeneity: {E}instein-{M}axwell-{S}calar {F}ield in {S}urface {S}ymmetry}, 
      author={W. Li},
      year={2024},
      eprint={2408.12434},
      archivePrefix={arXiv},
      primaryClass={gr-qc},
      url={https://arxiv.org/abs/2408.12434}, 
}

@article{Li:2024b,
      title={{BKL} {B}ounces {O}utside {H}omogeneity: {G}owdy {S}ymmetric {S}pacetimes}, 
      author={W. Li},
      year={2024},
      eprint={2408.12427},
      archivePrefix={arXiv},
      primaryClass={gr-qc},
      url={https://arxiv.org/abs/2408.12427}, 
}

@article{Chiodaroli:2014,
   title={{A} {C}ounterexample to {W}ell-{P}osedness of {E}ntropy {S}olutions to the {C}ompressible {E}uler {S}ystem},
   volume={11},
   ISSN={1793-6993},
   url={http://dx.doi.org/10.1142/S0219891614500143},
   DOI={10.1142/s0219891614500143},
   number={03},
   journal={Journal of Hyperbolic Differential Equations},
   publisher={World Scientific Pub Co Pte Lt},
   author={Chiodaroli, E.},
   year={2014},
   month=sep, pages={493–519} }

@article{Chiodaroli_et_al:2014,
   title={{G}lobal {I}ll‐{P}osedness of the {I}sentropic {S}ystem of {G}as {D}ynamics},
   volume={68},
   ISSN={1097-0312},
   url={http://dx.doi.org/10.1002/cpa.21537},
   DOI={10.1002/cpa.21537},
   number={7},
   journal={Communications on Pure and Applied Mathematics},
   publisher={Wiley},
   author={Chiodaroli, E. and De Lellis, C. and Kreml, O.},
   year={2014},
   month=aug, pages={1157–1190} }

@article{ColeyLim:2013,
   title={General {R}elativistic {D}ensity {P}erturbations},
   volume={31},
   ISSN={1361-6382},
   url={http://dx.doi.org/10.1088/0264-9381/31/1/015020},
   DOI={10.1088/0264-9381/31/1/015020},
   number={1},
   journal={Classical and Quantum Gravity},
   publisher={IOP Publishing},
   author={Lim, W.C. and Coley, A.A.},
   year={2013},
   month=nov, pages={015020} }

@article{ColeyLim:2012,
  title = {{G}enerating {M}atter {I}nhomogeneities in {G}eneral {R}elativity},
  author = {Coley, A.A. and Lim, W.C.},
  journal = {Phys. Rev. Lett.},
  volume = {108},
  issue = {19},
  pages = {191101},
  numpages = {5},
  year = {2012},
  publisher = {American Physical Society},
  doi = {10.1103/PhysRevLett.108.191101},
  url = {https://link.aps.org/doi/10.1103/PhysRevLett.108.191101}
}

@article{Oliynyk:2024,
  title={{O}n the {F}ractional {D}ensity {G}radient {B}low-{U}p {C}onjecture of {R}endall},
  author={Oliynyk, T.A.},
  journal={Communications in Mathematical Physics},
  volume={405},
  number={8},
  pages={197},
  year={2024},
  publisher={Springer}
}

@article{Hervik_et_al:2010,
  title={{F}uture {A}symptotics of {T}ilted {B}ianchi {T}ype {II} {C}osmologies},
  author={Hervik, S. and Lim, W.C. and Sandin, P. and Uggla, C.},
  journal={Class. Quantum Gravity},
  volume={27},
  number={18},
  pages={185006},
  year={2010},
  publisher={IOP Publishing}
}

@phdthesis{Lim_Thesis:2004,
    title={The {D}ynamics of {I}nhomogeneous {C}osmologies}, 
    author={W.C. Lim},
    year={2004},
    school = {University of {W}aterloo},
    eprint={gr-qc/0410126},
      archivePrefix={arXiv},
      primaryClass={gr-qc},
      url={https://arxiv.org/abs/gr-qc/0410126}
}

@article{BergerColella:1989,
title = {Local {A}daptive {M}esh {R}efinement for {S}hock {H}ydrodynamics},
journal = {Journal of Computational Physics},
volume = {82},
number = {1},
pages = {64-84},
year = {1989},
issn = {0021-9991},
doi = {https://doi.org/10.1016/0021-9991(89)90035-1},
author = {M.J. Berger and P. Colella},
}

@article{Ellis:1967,
  title={{D}ynamics of {P}ressure-{F}ree {M}atter in {G}eneral {R}elativity},
  author={Ellis, G.F.R.},
  journal={Journal of Mathematical Physics},
  volume={8},
  number={5},
  pages={1171--1194},
  year={1967},
  publisher={American Institute of Physics}
}

@article{EllisKing:1973,
	author = {King, A.R. and Ellis, G.F.R.},
	date = {1973/09/01},
	doi = {10.1007/BF01646266},
	id = {King1973},
	isbn = {1432-0916},
	journal = {Communications in Mathematical Physics},
	number = {3},
	pages = {209--242},
	title = {{T}ilted {H}omogeneous {C}osmological {M}odels},
	volume = {31},
	year = {1973}}

@article{WainwrightHsu:1989,
year = {1989},
month = {oct},
volume = {6},
number = {10},
pages = {1409},
author = {Wainwright, J. and Hsu,  L.},
title = {{A} {D}ynamical {S}ystems {A}pproach to {B}ianchi {C}osmologies: {O}rthogonal {M}odels of {C}lass {A}},
journal = {Classical and Quantum Gravity}
}

@article{DiazKurganovdeLuna:2019,
  title={{P}ath-{C}onservative {C}entral-{U}pwind {S}chemes for {N}onconservative {H}yperbolic {S}ystems},
  author={Diaz, M.J.C. and Kurganov, A. and de Luna, T.M.},
  journal={ESAIM: Mathematical Modelling and Numerical Analysis},
  volume={53},
  number={3},
  pages={959--985},
  year={2019},
  publisher={EDP Sciences}
}

@article{Brehm:2016,
title={{B}ianchi {VIII} and {IX} {V}acuum {C}osmologies: {A}lmost {E}very {S}olution {F}orms {P}article {H}orizons and {C}onverges to the {M}ixmaster {A}ttractor},
author={Bernhard Brehm},
year={2016},
eprint={1606.08058},
archivePrefix={arXiv},
url={https://arxiv.org/abs/1606.08058}, 
}

@article{HeinzleUggla:2009,
  title={A {N}ew {P}roof of the {B}ianchi {T}ype {IX} {A}ttractor {T}heorem},
  author={Heinzle, J.M and Uggla, C.},
  journal={Class. Quantum Grav.},
  volume={26},
  number={7},
  pages={075015},
  year={2009},
  publisher={IOP Publishing}
}

@article{Pares:2006,
  title={Numerical {M}ethods for {N}onconservative {H}yperbolic {S}ystems: {A} {T}heoretical {F}ramework.},
  author={Par{\'e}s, C.},
  journal={SIAM Journal on Numerical Analysis},
  volume={44},
  number={1},
  pages={300--321},
  year={2006},
  publisher={SIAM}
}

@article{DLM:1995,
  title={Definition and {W}eak {S}tability of {N}onconservative {P}roducts},
  author={Dal Maso, G. and LeFloch, P.G. and Murat, F.},
  journal={Journal de math{\'e}matiques pures et appliqu{\'e}es},
  volume={74},
  number={6},
  pages={483--548},
  year={1995},
  publisher={Elsevier}
}

@article{FMO:2025,
      title={{F}uture {S}tability of {T}ilted {T}wo-{F}luid {B}ianchi {I} {S}pacetimes}, 
      author={G. Fournodavlos and E. Marshall and T.A. Oliynyk},
      year={2025},
      eprint={2508.15155},
      archivePrefix={arXiv},
      primaryClass={gr-qc},
      url={https://arxiv.org/abs/2508.15155}, 
}

@article{BeyerOliynyk:2024b,
   title={Relativistic {P}erfect {F}luids {N}ear {K}asner {S}ingularities},
   volume={32},
   ISSN={1944-9992},
   url={http://dx.doi.org/10.4310/CAG.241204004223},
   DOI={10.4310/cag.241204004223},
   number={6},
   journal={Communications in Analysis and Geometry},
   publisher={International Press of Boston},
   author={Beyer, F. and Oliynyk, T.A.},
   year={2024},
   pages={1701–1794} }

@article{Castro_et_al:2013,
  title={Entropy {C}onservative and {E}ntropy {S}table {S}chemes for {N}onconservative {H}yperbolic {S}ystems},
  author={Castro, M.J. and Fjordholm, U.S. and Mishra, S. and Par{\'e}s, C.},
  journal={SIAM Journal on Numerical Analysis},
  volume={51},
  number={3},
  pages={1371--1391},
  year={2013},
  publisher={SIAM}
}

@article{AbgrallKarni:2010,
  title={A {C}omment on the {C}omputation of {N}on-{C}onservative {P}products},
  author={Abgrall, R. and Karni, S.},
  journal={Journal of Computational Physics},
  volume={229},
  number={8},
  pages={2759--2763},
  year={2010},
  publisher={Elsevier}
}

@article{Castro_et_al:2008,
  title={Why {M}any {T}heories of {S}hock {W}aves are {N}ecessary: {C}onvergence {E}rror in {F}ormally {P}ath-{C}onsistent {S}chemes},
  author={Castro, M.J. and LeFloch, P.G. and Mu{\~n}oz-Ruiz, M.L. and Par{\'e}s, C.},
  journal={Journal of Computational Physics},
  volume={227},
  number={17},
  pages={8107--8129},
  year={2008},
  publisher={Elsevier}
}

@article{ZhangShu:2011,
  title={Positivity-{P}reserving {H}igh {O}rder {D}iscontinuous {G}alerkin {S}chemes for {C}ompressible {E}uler {E}quations with {S}ource {T}erms},
  author={Zhang, X. and Shu, C.},
  journal={Journal of Computational Physics},
  volume={230},
  number={4},
  pages={1238--1248},
  year={2011},
  publisher={Elsevier}
}

@article{ZhangShu:2010,
  title={On {P}ositivity-{P}reserving {H}igh {O}rder {D}iscontinuous {G}alerkin {S}chemes for {C}ompressible {E}uler {E}quations on {R}ectangular {M}eshes},
  author={Zhang, X. and Shu, C.},
  journal={Journal of Computational Physics},
  volume={229},
  number={23},
  pages={8918--8934},
  year={2010},
  publisher={Elsevier}
}

@article{Wu:2017,
  title={Design of {P}rovably {P}hysical-{C}onstraint-{P}reserving {M}ethods for {G}eneral {R}elativistic {H}ydrodynamics},
  author={Wu, K.},
  journal={Physical Review D},
  volume={95},
  number={10},
  pages={103001},
  year={2017},
  publisher={APS}
}

@article{ColeyLim:2023,
  title={Periodic {B}oundary {C}onditions and {G}2 {C}osmology},
  author={Coley, A.A. and Lim, W.C.},
  journal={Classical and Quantum Gravity},
  volume={41},
  number={1},
  pages={015009},
  year={2023},
  publisher={IOP Publishing}
}

@article{Garfinkle_et_al:2023,
  title={Initial {C}onditions {P}roblem in {C}osmological {I}nflation {R}evisited},
  author={Garfinkle, D. and Ijjas, A. and Steinhardt, P.J.},
  journal={Physics Letters B},
  volume={843},
  pages={138028},
  year={2023},
  publisher={Elsevier}
}

@article{Aurrekoetxea_et_al:2023,
   title={{CTTK}: {A} {N}ew {M}ethod to {S}olve the {I}nitial {D}ata {C}onstraints in {N}umerical {R}elativity},
   volume={40},
   ISSN={1361-6382},
   url={http://dx.doi.org/10.1088/1361-6382/acb883},
   DOI={10.1088/1361-6382/acb883},
   number={7},
   journal={Classical and Quantum Gravity},
   publisher={IOP Publishing},
   author={Aurrekoetxea, J.C. and Clough, K. and Lim, E.A.},
   year={2023},
   month=mar, pages={075003} }

@article{Racz_et_al:2021,
   title={The {A}nisotropy of the {P}ower {S}pectrum in {P}eriodic {C}osmological {S}imulations},
   volume={503},
   ISSN={1365-2966},
   url={http://dx.doi.org/10.1093/mnras/stab874},
   DOI={10.1093/mnras/stab874},
   number={4},
   journal={Monthly Notices of the Royal Astronomical Society},
   publisher={Oxford University Press (OUP)},
   author={Rácz, G. and Szapudi, I. and Csabai, I. and Dobos, L.},
   year={2021},
month=mar, pages={5638–5645} }

@article{Coley_et_al:2006b,
  title={Tilt and {P}hantom {C}osmology},
  author={Coley, A.A. and Hervik, S. and Lim, W.C.},
  journal={Physics Letters B},
  volume={638},
  number={4},
  pages={310--313},
  year={2006},
  publisher={Elsevier}
}

@article{ColeyHervik:2005,
  title={A {D}ynamical {S}ystems {A}pproach to the {T}ilted {B}ianchi {M}odels of {S}olvable {T}ype},
  author={Coley, A. and Hervik, S.},
  journal={Classical and Quantum Gravity},
  volume={22},
  number={3},
  pages={579},
  year={2005},
  publisher={IOP Publishing}
}

@article{ColeyHervik:2004,
  title={Bianchi {C}osmologies: {A} {T}ale of {T}wo {T}ilted {F}luids},
  author={Coley, A.A. and Hervik, S.},
  journal={Classical and Quantum Gravity},
  volume={21},
  number={17},
  pages={4193},
  year={2004},
  publisher={IOP Publishing}
}

@article{Hervik_et_al:2006,
  title={The {F}utures of {B}ianchi {T}ype {VII0} {C}osmologies with {V}orticity},
  author={Hervik, S. and Van Den Hoogen, R.J. and Lim, W.C. and Coley, A.A.},
  journal={Classical and Quantum Gravity},
  volume={23},
  number={3},
  pages={845},
  year={2006},
  publisher={IOP Publishing}
}

@article{Coley_et_al:2006a,
  title={Fluid {O}bservers and {T}ilting {C}osmology},
  author={Coley, A.A. and Hervik, S. and Lim, W.C.},
  journal={Classical and Quantum Gravity},
  volume={23},
  number={10},
  pages={3573},
  year={2006},
  publisher={IOP Publishing}
}

@article{Hervik_et_al:2007a,
  title={Late-{T}ime {B}ehaviour of the {T}ilted {B}ianchi {T}ype {$VI_{h}$} {M}odels},
  author={Hervik, S. and Van den Hoogen, R.J. and Lim, W.C. and Coley, A.A.},
  journal={Classical and Quantum Gravity},
  volume={24},
  number={15},
  pages={3859},
  year={2007},
  publisher={IOP Publishing}
}

@article{Hervik_et_al:2007b,
  title={Late-{T}ime {B}ehaviour of the {T}ilted {B}ianchi {T}ype {$VI_{-1/9}$} {M}odels},
  author={Hervik, S. and van den Hoogen, R.J. and Lim, W.C. and Coley, A.A.},
  journal={Classical and Quantum Gravity},
  volume={25},
  number={1},
  pages={015002},
  year={2007},
  publisher={IOP Publishing}
}

@article{ColeyHervik:2008,
  title={Bianchi {M}odels with {V}orticity: {T}he {T}ype {III} {B}ifurcation},
  author={Coley, A. and Hervik, S.},
  journal={arXiv preprint arXiv:0802.3629},
  year={2008}
}

\end{document}